%% file: main.tex
\documentclass[a4paper,11pt]{article}
\pdfoutput=1 

\usepackage{jheppub} 
\usepackage{float}
\usepackage[OT1]{fontenc} 
\usepackage{subcaption}

\title{\boldmath TRANSIT your events into a new mass: Fast background interpolation for weakly-supervised anomaly searches
}

\author[ab]{I. Oleksiyuk,}
\author[b]{S. Voloshynovskiy,}
\author[a]{T. Golling}

\affiliation[a]{Département de Physique Nucléaire et Corpusculaire, University of Geneva, 1211 Geneva, Switzerland}
\affiliation[b]{Department of Computer Science, University of Geneva,
 Route de Drize 7, 1211 Geneva, Switzerland}

\emailAdd{ivan.oleksiyuk@unige.ch}
\emailAdd{svyatoslav.voloshynovskyy@unige.ch}
\emailAdd{tobias.golling@unige.ch}

\abstract{
We introduce a new model for conditional and continuous data morphing called TRansport Adversarial Network for Smooth InTerpolation (TRANSIT). We apply it to create a background data template for weakly-supervised searches at the LHC. The method smoothly transforms sideband events to match signal region mass distributions.
We demonstrate the performance of TRANSIT using the LHC Olympics R\&D dataset. The model captures non-linear mass correlations of features and produces a template that offers a competitive anomaly sensitivity compared to state-of-the-art transport-based template generators. Moreover, the computational training time required for TRANSIT is an order of magnitude lower than that of competing deep learning methods. This makes it ideal for analyses that iterate over many signal regions and signal models.
Unlike generative models, which must learn a full probability density distribution, i.e., the correlations between all the variables, the proposed transport model only has to learn a smooth conditional shift of the distribution. This allows for a simpler, more efficient residual architecture, enabling mass uncorrelated features to pass the network unchanged while the mass correlated features are adjusted accordingly. Furthermore, we show that the latent space of the model provides a set of mass decorrelated features useful for anomaly detection without background sculpting.
}

\begin{document} 
\maketitle
\flushbottom

\section{Introduction}
\input{tex/intro}

\section{Dataset}
\input{tex/data}

\section{Method}
\input{tex/method}

\section{Results}
\input{tex/results}

\section{Conclusions}
\input{tex/conclusion}

\acknowledgments

The authors would like to acknowledge funding through the SNSF Sinergia grant CRSII5\_193716 ``Robust Deep Density Models for High-Energy Particle Physics and Solar Flare Analysis (RODEM)''
and the SNSF project grant 200020\_212127 ``At the two upgrade frontiers: machine learning and the ITk Pixel detector''.

\section*{Code availability}

The code used to produce all results presented
in this work is available publicly at \url{https://github.com/IvanOleksiyuk/transit-hep} .

\section*{Appendix}
\appendix
\input{tex/app}

\bibliography{bibliography}
\bibliographystyle{JHEP}

\end{document}

%% file: tex/intro.tex
\label{sec:intro} 
Since the discovery of the Higgs boson in 2012 \cite{Aad_2012, Chatrchyan_2012}, the Standard Model (SM) of particle physics has shown phenomenal agreement with most experimental data collected at the Large Hadron Collider (LHC).
Despite its success, the SM still fails in explaining gravity, neutrino masses, and dark matter, among other shortcomings.
The majority of Beyond Standard Model (BSM) theories assume the existence of yet-undiscovered particles, motivating the searches for resonances in the spectra of the invariant mass. 
However, the proposed particles differ from the known ones not only in their mass but also in many other observables. 
Selecting events with a specific model-dependent signature greatly increases the sensitivity of a search to that model's signal, but, in general, becomes less sensitive to other signals.
Despite this, scanning the parameter space of all the proposed BSM models with model-specific searches is extremely resource-consuming.
Moreover, the actual BSM physics might lie outside the scope of current theoretical proposals.
To address this issue, numerous machine learning (ML) methods capable of detecting a wide range of signals have been developed
\cite{CWOLA,SALAD,FETA,CATHODE,CURTAINS, CURTAINSF4F,DRAPES,FullPhase,SIGMA,Zhang:2024ebl,Chekanov:2024ezm,Matos:2024ggs,Grosso:2024nho,Li:2024htp,Oleksiyuk:2024hru,Cheng:2024yig,Krause:2023uww,Zipper:2023ybp,Metodiev:2023izu,Liu:2023djx,Zhang:2023khv,Bai:2023yyy,Freytsis:2023cjr,Finke:2023ltw,Bickendorf:2023nej,Chekanov:2023uot,ATLAS:2023azi,Vaslin:2023lig,Golling:2023yjq,Mikuni:2023tok,Golling:2023juz,Roche:2023int,Schuhmacher:2023pro,Mastandrea:2022vas,Araz:2022zxk,Kasieczka:2022naq,Kamenik:2022qxs,Park:2022zov,Caron:2022wrw,Dillon:2022mkq,Verheyen:2022tov,Finke:2022lsu,Fanelli:2022xwl,Letizia:2022xbe,Birman:2022xzu,Dillon:2022tmm,Jiang:2022sfw,Alvi:2022fkk,Buss:2022lxw,Aguilar-Saavedra:2022ejy,Bradshaw:2022qev,Ngairangbam:2021yma,Canelli:2021aps,dAgnolo:2021aun,Chekanov:2021pus,Mikuni:2021nwn,Lester:2021aks,Tombs:2021wae,Aguilar-Saavedra:2021utu,Herrero-Garcia:2021goa,Jawahar:2021vyu,Fraser:2021lxm,Ostdiek:2021bem,Govorkova:2021utb,Volkovich:2021txe,Kasieczka:2021tew,Govorkova:2021hqu,Caron:2021wmq,Dorigo:2021iyy,Aarrestad:2021oeb,Kahn:2021drv,Atkinson:2021nlt,Shih:2021kbt,Finke:2021sdf,Dillon:2021nxw,Collins:2021nxn,Bortolato:2021zic,Blance:2021gcs,Batson:2021agz,Chakravarti:2021svb,Kasieczka:2021xcg,Stein:2020rou,Faroughy:2020gas,Park:2020pak,vanBeekveld:2020txa,Mikuni:2020qds,pol2020anomaly,1815227,aguilarsaavedra2020mass,Alexander:2020mbx,Thaprasop:2020mzp,Khosa:2020qrz,Cheng:2020dal,Amram:2020ykb,1800445,1797846,knapp2020adversarially,Romao:2020ojy,Romao:2019dvs,Aguilar-Saavedra:2017rzt,Nachman:2020lpy,Dillon:2019cqt,1809.02977,Mullin:2019mmh,DeSimone:2018efk,Hajer:2018kqm,Blance:2019ibf,Cerri:2018anq,Roy:2019jae,Heimel:2018mkt,Farina:2018fyg,DAgnolo:2019vbw,Collins:2019jip,Collins:2018epr,DAgnolo:2018cun}. Several works already apply these methods to real data in high-energy physics (HEP) analyses at ATLAS \cite{collaboration2020dijet, atlascollaboration2025weaklysupervisedanomalydetection} CMS \cite{CMS:2024nsz} and DARWIN \cite{DARWIN:2024unx}.

A prominent class of model-agnostic methods is weakly supervised anomaly search, which was first introduced to HEP as Classification Without Labels (CWoLa) \cite{CWOLA}.
This and many of the subsequent methods \cite{SALAD, FETA, CURTAINS, CURTAINSF4F, CATHODE, DRAPES, FullPhase, SIGMA} can be described by the same algorithm.
First, a signal region (SR) is selected, i.e., a window in the distribution of the resonant variable $m$, where the signal peak is supposedly localised. The rest of the resonant variable spectrum is then assumed to be nearly signal-free. A part of this signal-poor region is, usually called sidebands (SB), is used to estimate the distribution of the additional observables $\boldsymbol{x}$ for the background data $p^{background}(\boldsymbol{x}|m)\approx p^{\text{data}}(\boldsymbol{x}|m) \approx p_{\Theta}(\boldsymbol{x}|m)\text{ for } m \in \text{SB}$ using parametrised models with parameters $\Theta$.
This distribution is then interpolated from $m \in \text{SB}$ to $m \in \text{SR}$ to produce a signal-poor template $p_{\Theta}(\boldsymbol{x}|m) \text{ for } m \in \text{SR}$. Finally, a classifier is trained based on observables $\boldsymbol{x}$ to distinguish between the signal-poor template and the signal-rich SR.
The pivotal point of these approaches is to find a method for high-quality template generation, as a poor-quality template will lead to a high false-positive rate of the CWoLa classifier. 
The original CWoLa implementation \cite{CWOLA} suggests taking the sideband data itself as a crude approximation of the background in SR.
A better template can be provided by Monte Carlo generation with reweighting using SALAD \cite{SALAD} or corrections using FETA \cite{FETA} methods, but it is more desirable to have a fully data-driven method due to the limited availability of high-quality simulation. To the best of our knowledge, all state-of-the-art (SotA) data-driven DL methods \cite{CURTAINS, CURTAINSF4F, CATHODE, DRAPES, FullPhase, SIGMA} rely on either normalising flows, diffusion, or a mixture of the two, such as continuous normalising flows (CNF). These methods provide high-quality templates but at a high computational cost, requiring hours to train even on relatively small datasets.

Despite the apparent simplicity of the semi-supervised framework, it usually results in a computationally demanding analysis for several reasons. First of all, location of a supposed signal mass peak is unknown. Thus, one has to apply this method over an order of 10 mass windows.
Before unblinding the experimental data, the method at hand has to be rigorously validated by applying it on tens of validation datasets and iterating over tens of random seeds to properly assess the uncertainty arising from the stochastic nature of the DL model fit.
Despite such analysis being model-agnostic, it also makes sense to assess the sensitivity of the analysis to tens of different signal models by injecting varying quantities of each signal.
This will also help set the limits on the BSM models in case the analysis shows no significant signal presence. 
Considering that these factors are multiplicative with one another, the typical HEP analysis would have to run this pipeline thousands of times, translating into extreme computational cost.
This presents a need for orders-of-magnitude improvement in method speed and efficiency, which has become the topic of the most recent studies. Two methods, namely CURTAINsF4F \cite{CURTAINSF4F} and SIGMA \cite{SIGMA}, investigated efficient ways to reuse the model trained on the entire mass spectrum in every signal region. 
Additionally, RAD-OT \cite{RADOT} exploits a non-ML-based optimal transport prescription to interpolate the sidebands in the signal region, trading a reduction in template generation time for lower template quality.

In this work, we address the issue of fast generation of high-quality templates by introducing the TRansport Adversarial Network for Smooth InTerpolation (TRANSIT). To increase efficiency, the method leverages the strategy of transporting data from sidebands into the SR, as in CURTAINs \cite{CURTAINS} and RAD-OT \cite{RADOT}, rather than generating the samples from noise. Moreover, the speed-up is achieved thanks to the simplicity of the network's one-pass feed-forward architecture, which requires less training time than most flow- and diffusion-based methods.
At the same time, it provides a template of quality competitive with other methods by employing specifically designed losses. As an additional benefit, the chosen losses lead to independence of the latent space variables from the resonant mass, allowing for an approach to mitigate background sculpting similar to LaCATHODE \cite{LaCATHODE}.

The remainder of this paper is organised as follows: Section \;\ref{sec:data} briefly describes the LHC Olympics (LHCO) R\&D dataset, which is used for the comparison of methods. Section\;\ref{sec:method} introduces the TRANSIT method and explains its working principle.
Subsequently, Section \;\ref{sec:results} presents the performance of the TRANSIT method, comparing it with other approaches.
Finally, Section\;\ref{sec:conclusion} provides the conclusions and outlook.

%% file: tex/data.tex
\label{sec:data}

One of the most suitable places to apply anomaly detection is a dijet BSM search.
Firstly, the hadronic dijet final state is a common signature in high-energy proton-proton collisions. Due to the high background of QCD jets, such a search could greatly benefit from anomaly detection methods aimed at enhancing signal significance.
Secondly, BSM signals can produce a variety of unusual jet substructures, e.g., semi-visible jets \cite{Cohen:2015toa}, emerging jets \cite{Schwaller:2015gea}, and 4-prong jets \cite{Schwaller:2015gea}, so a model-unspecific method is preferred.

The LHCO R\&D \cite{LHCO} dataset consists of 1 million background dijet events from SM quark/gluon scattering and 100 thousand signal dijet events produced through a BSM resonance $Z'\rightarrow X(\rightarrow qq)Y(\rightarrow qq)$ events.
The resonance has a mass $m_Z = 3.5$ TeV, and the decay products have asymmetric masses $m_X = 500$ GeV and $m_Y = 100$ GeV.
The dataset is simulated using \textsc{Pythia}~8.219~\cite{Sjostrand:2014zea} and \textsc{Delphes}~3.4.1~\cite{de_Favereau_2014, Mertens_2015, Selvaggi_2014} with default settings.
Jets are clustered using the anti-$k_T$ algorithm with radius $R=1.0$, implemented in the \textsc{FastJet} package \cite{Cacciari:2011ma}.
Only events that have at least one jet with transverse momentum $p^J_T > 1.2$ TeV and pseudorapidity $\eta < 2.5$ are kept. In each event, only the two leading jets are retained.

\begin{figure}[H]
    \centering
    \includegraphics[width=1\linewidth]{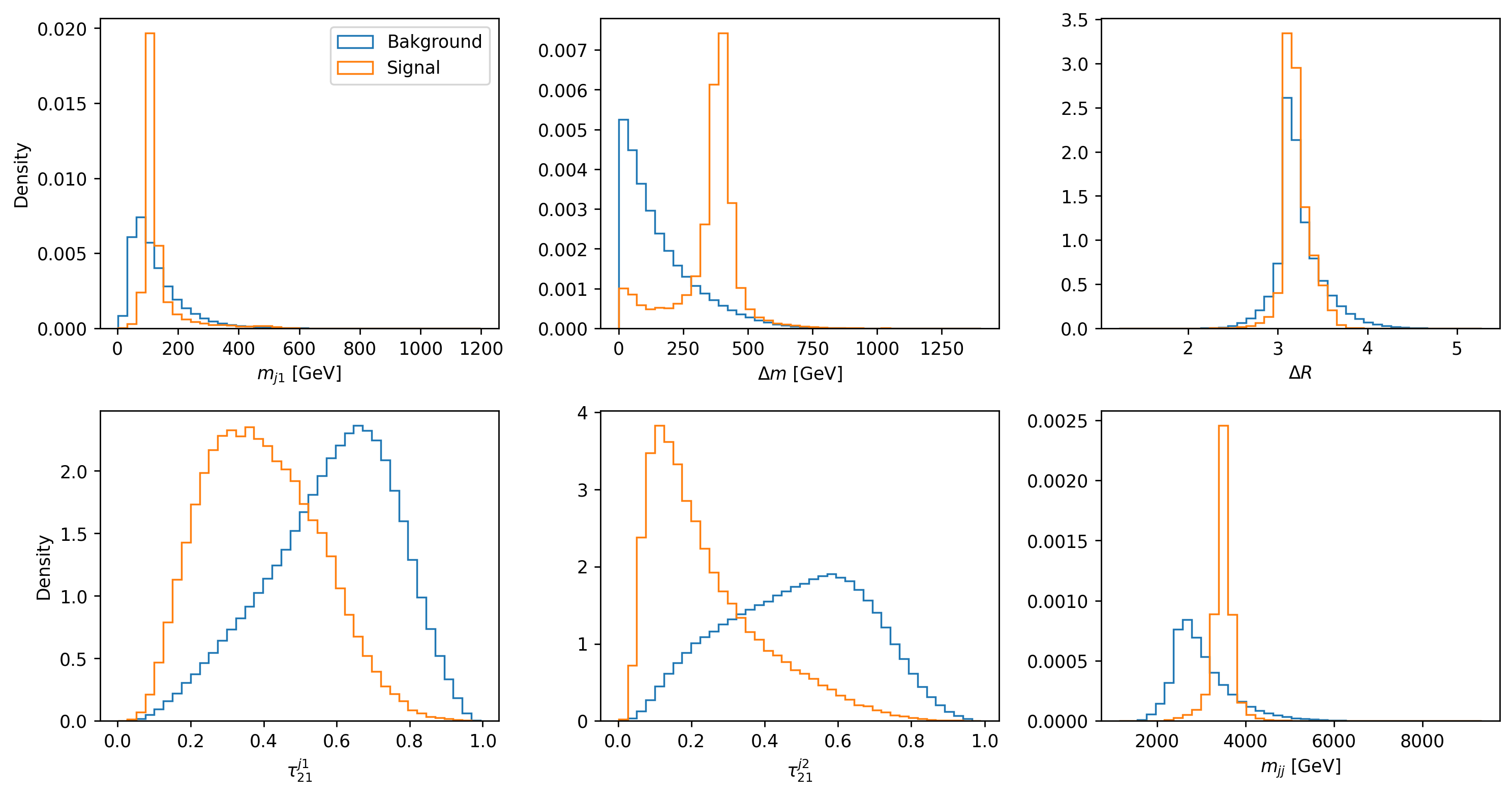}
    \caption{Distributions of high-level observables commonly used in weakly supervised searches within the LHCO R\&D dataset, presented for the QCD background and $Z'$ signal.}
    \label{fig:LHCO}
\end{figure}

In order to compare with existing template generation methods, we apply TRANSIT to a commonly used set of high-level variables for dijet events: the mass of the heavier leading jet $m_{j1}$, the mass difference between the two leading jets $\Delta m$, the distance $\Delta R$ between these jets in $(\phi, \eta)$ space, and the two-to-one subjetiness ratios $\tau^{j1}_{2,1}$ and $\tau^{j2}_{2,1}$. The distributions of the selected observables are shown in Fig. \ref{fig:LHCO}.
In addition, we select the interval $[3.3, 3.7]$ TeV as the signal region and $[3.0, 3.3]$ TeV and $[3.7, 4.6]$ TeV as the sideband regions.
For evaluation and training, we use all the background available in these regions, but we also add $N_{\text{sig}}$ if signal contamination is required.
The signal events are sampled randomly based on the training seed, so their stochasticity is included in the errorbars on the plots in Section \ref{sec:results}.

%% file: tex/method.tex
\label{sec:method}

\subsection{Main principles}
In general, generative models aim at approximating the joint probability distribution of observables $\boldsymbol{X} = (X_1, \ldots, X_n)$ conditioned on the mass $M$, namely $p(X_1, X_2, \ldots, X_n | M)$ \footnote{In our notation, $p(\boldsymbol{X})\equiv p_{\boldsymbol{X}}$ refers to the probability distribution of a random vector $\boldsymbol{X}$ as a function, while $p(\boldsymbol{x}) \equiv p_{\boldsymbol{X}}(\boldsymbol{x}) \equiv p(\boldsymbol{X}=\boldsymbol{x})$ denotes the value of this function for a specific sample $\boldsymbol{x}$.}.
Examples of such approaches in the weakly-supervised search context include CATHODE \cite{CATHODE}, DRAPES \cite{DRAPES}, and others \cite{FullPhase, SIGMA}. 
The advantage of these methods is that, after training, one can sample events $\boldsymbol{x} \sim p(\boldsymbol{X}|M)$ at will for any mass. 
However, the model must learn not only the correlations between each of the variables $X_1, \ldots, X_n$ and the mass $M$, as well as the mass-dependent correlations among the variables $X_1, \ldots, X_n$, but also the mass-independent correlations among these variables.

As an alternative, one can train a transport model represented by the functional form $f(\boldsymbol{x}, m, \hat{m}) \xrightarrow{} \boldsymbol{\hat{x}}$ that would transform samples of an original mass $\boldsymbol{x} \sim p(\boldsymbol{X}|M=m)$ into samples corresponding to a new target mass $\boldsymbol{\hat{x}} \sim p(\boldsymbol{X}|M=\hat{m})$.
If a variable $X_k$ is uncorrelated with the mass $M$, the model can satisfy this condition by simply learning the identity transformation.
The same holds for the correlations between variables $X_i$ and $X_j$. If the correlation does not change with the mass, i.e., if one can achieve the correct conditional distribution by transporting each variable separately, then there is no need to learn the correlation between them.
For variables with smooth mass dependence, the method would have to simply learn a correction shift to transport events along smooth trajectories into a different mass, as illustrated in the right part of Fig.\,\ref{fig:losses}.
This reduces the total amount of correlations that have to be encoded in the network compared to a full-generation case, so the transport network should require fewer parameters and less training time.
We provide an extended argumentation in App.\,\ref{app:transport_vs_generation}.

In the literature, this approach was introduced with the method CURTAINs \cite{CURTAINS}, which is based on training an invertible neural network (INN) conditioned on both the original and target mass.
The challenges of estimating $p(\boldsymbol{X}|M)$ for INN optimisation and the computational expense of training led to the development of an extension of the method in CURTAINs Flows for Flows (F4F) \cite{CURTAINSF4F}, which achieved SotA performance at the time. 
However, the method remains rather computationally demanding.
A more recent method, RAD-OT \cite{RADOT}, uses optimal transport to interpolate the template between two sidebands.
Despite being computationally light, the method has limited template-building quality, as it only offers a linear interpolation path for each event, neglecting the apparent trends in the sideband regions.
These two methods are thus closely related to TRANSIT and will be used for benchmarking.

A different perspective on the template generation problem was introduced in LaCATHODE \cite{LaCATHODE} by prioritising background sculpting mitigation.
A conditional normalising flow is used to provide mass-decorrelated variables $\boldsymbol{z}$ in the latent space, which is restricted to have a multivariate unit Gaussian distribution.
The variables $\boldsymbol{z}$ are then used as a basis for sculpting-free CWoLa-style analysis.
Despite this transformation being sufficient for decorrelation, it is excessively restrictive, as it is only necessary that the latent distribution does not depend on mass.
For example, in cases where the input variables are already mass-decorrelated, there is no need to transform them into a Gaussian.

In this work, we show that non-linear smooth transport and latent mass decorrelation can both be achieved simultaneously by training a simple residual multi-layer perceptron (MLP) that is efficiently parallelisable on modern hardware, thus leading to significant speedups.
The remaining challenge is to design a set of loss functions that satisfy the transport and decorrelation objectives.

\subsection{TRANSIT model}
\label{subsec:transit_model}
\begin{figure}[h]
    \centering
    \includegraphics[width=1\linewidth]{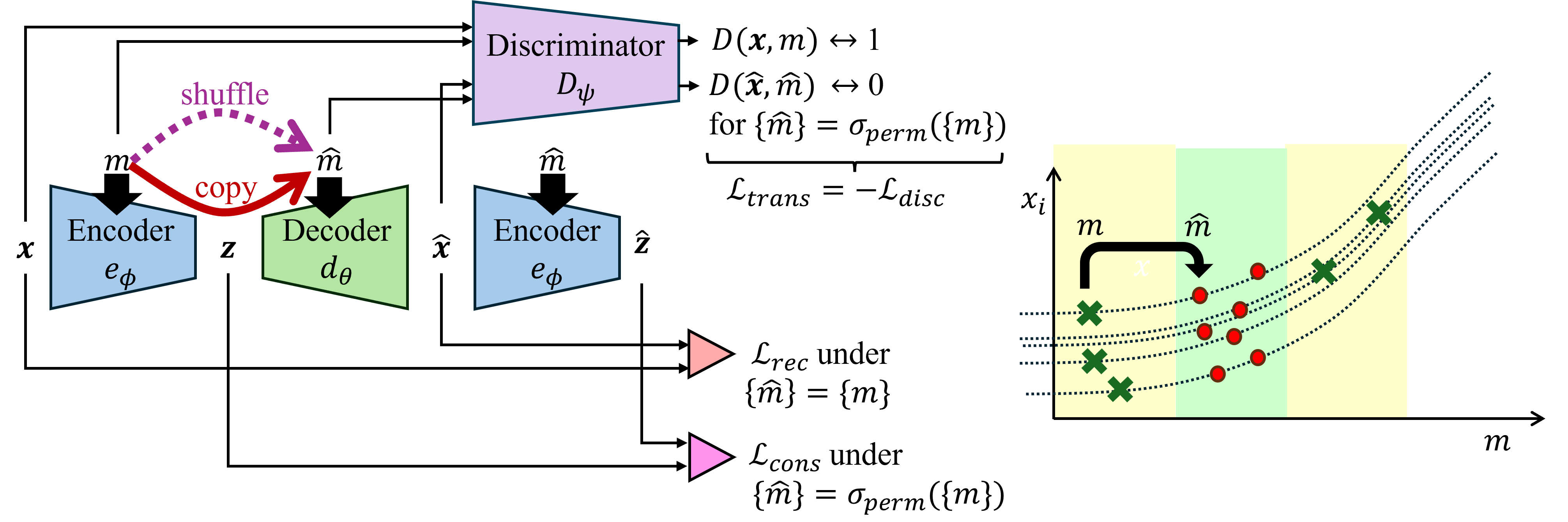}
    \caption{The transport of event form mass $m$ into mass $\hat{m}$. Left: Passage of data through the TRANSIT model with all the main components and losses. Right: The principle of transporting original events in sidebands (green crosses), corresponding to the original mass \(m\), along the transport curves (dotted lines) to transformed events (red circles) corresponding to a new mass \(\hat{m}\) in the signal region. $\sigma_{perm}$ denotes an operation of random permutation of the batch.}
    \label{fig:losses}
\end{figure}

The TRANSIT model consists of several key components, depicted on the left in Fig.\,\ref{fig:losses}. 
Starting with the true data event pair $(\boldsymbol{x}, m)$, the model passes each event $\boldsymbol{x}$ via the encoder network $e_\phi$ conditioned on the corresponding mass $m$, so that it is encoded in the latent representation $\boldsymbol{z} = e_\phi(\boldsymbol{x}, m)$. 
The dimensionality of the latent space, \(D_{\boldsymbol{z}}\), may differ from that of the input space, \(D_{\boldsymbol{x}}\); we choose \(D_{\boldsymbol{z}} > D_{\boldsymbol{x}}\) to reduce information loss within the network.
After that, the model decodes the latent representation conditionally on the target mass $\hat{m}$ into an event $\boldsymbol{\hat{x}} = d_\theta(\boldsymbol{z}, \hat{m})$ of the same dimensionality as $\boldsymbol{x}$. The mass $\hat{m}$ can be equal to or different from $m$ depending on the context.
The aim of the encoder is to decorrelate the variables $\boldsymbol{x}$ from the original mass $m$, so that one obtains a mass-independent latent representation $\boldsymbol{z}$, and then restores the correlation in the decoder with a target mass $\hat{m}$.
Together, the encoder and decoder form a transport model (TM), denoted as $f_{\phi, \theta}(\boldsymbol{x}, m, \hat{m})=d_\theta(e_\phi(\boldsymbol{x}, m), \hat{m})$.

\paragraph{Reconstruction loss.} 
In the case where $m=\hat{m}$, the event should be reconstructed as itself as in the case of auto-encoder.
To enforce this, the model passes a batch of $N$ events from SB $\{\boldsymbol{x}\}=\{\boldsymbol{x}^1 \ldots \boldsymbol{x}^N\}$ through TM with conditioning on the paired masses $\{m\}=\{m^1 \ldots m^N\}$ in both encoder and decoder (i.e. $\hat{m}^k=m^k$) and evaluate the mean squared error discrepancy between the input and output, averaged over the batch
\begin{equation}
\mathcal{L}_{\text{rec}} = \mathbb{E}_{(\boldsymbol{x}, m)\sim p_{data}(\boldsymbol{X}, M)}||\boldsymbol{x} - d_\theta(e_\phi(\boldsymbol{x}, m), m)||^2.
\end{equation}
During training, we first pre-train the network for several epochs to achieve a small reconstruction loss before enabling the rest of the losses discussed below.

\paragraph{Transport loss.} To train the transport into a new mass $\hat{m}$, we pass a randomly permuted (shuffled) batch of masses $\{\hat{m}\}=\sigma_{\text{perm}}(\{m\})$ to the conditional decoder. We require that the transported events created with these shuffled masses $\boldsymbol{\hat{x}}=d_\theta(e_\phi(\boldsymbol{x}, m), \hat{m})\sim \hat{p}_{\phi, \theta}(\hat{X}|\hat{M})$ follow the distribution of the events in the data $p_{\text{data}}(X|M)$. 
As the marginal distributions of masses in the batches are the same, $p(\hat{M})=p(M)$, using Bayes' theorem, we can conclude that
\begin{equation}
 \hat{p}_{\phi, \theta}(\hat{X}|\hat{M})=p_{\text{data}}(X|M)  \stackrel{p(M)=p(\hat{M})}{\Longleftrightarrow} \hat{p}_{\phi, \theta}(\hat{X}, \hat{M})=p_{\text{data}}(X, M).
\end{equation}
Thus, we have to minimise the discrepancy between these joint distributions.
In theory, these distributions can be compared using the Jensen-Shannon Divergence $JSD(\hat{p}_{\phi, \theta}||p_{\text{data}})$; however, it is usually computationally intractable. 
Instead, we use a density ratio estimation trick described in \cite{Goodfellow:2014upx}, which provides the basis for all generative adversarial networks. 
For an optimal discriminator $D$, the binary cross entropy loss is proportional to $JSD$ with an added constant 
\begin{equation}
\begin{split}
   &  BCE_D(p_{\text{data}}||\hat{p}_{\phi, \theta}) = -\mathbb{E}_{(\boldsymbol{x}, m)\sim p_{data}}[\ln(D(\boldsymbol{x}, m))]-\mathbb{E}_{(\boldsymbol{\hat{x}}, \hat{m})\sim\hat{p}_{\phi, \theta}}[\ln(1-D(\boldsymbol{\hat{x}}, \hat{m}))]\\
   & =\ln(4)-2 JSD(p_{\text{data}}||\hat{p}_{\phi, \theta}).
\end{split}
\label{eq:ldisc}
\end{equation}
This value can be approximated by training a parametrised binary classifier $D_\psi$ in place of an optimal classifier $D$ to distinguish between transported and true pairs (see Fig.\;\ref{fig:losses}), namely by minimising 
\begin{equation}
\mathcal{L}_{\text{disc}}=BCE_{D_\psi}(p_{\text{data}}||\hat{p}_{\phi, \theta})
\end{equation}
with respect to $\psi$, and continuously updating the classifier so that it remains close to optimal. 
Then, by maximising $\mathcal{L}_{\text{disc}}$ with respect to TM parameters $\phi, \theta$, i.e., ``fooling" the discriminator by creating more realistic samples, we can minimise $JSD$ between the generated and true distributions. 

In our particular case, we use a simple conditional multilayer perceptron (MLP) as the classifier. We optimise the classifier by performing steps in the $-\nabla_\psi \mathcal{L}_{\text{disc}}$ direction and use $\nabla_{\phi,\theta} \mathcal{L}_{\text{disc}}$ to update the TM.  The loss of the discriminator provides a meaningful step for the TM only if the discriminator is currently able to distinguish between the two distributions with the correct labels. Therefore, if $\mathcal{L}_{\text{disc}}>\ln(4)$, only the discriminator training steps are performed, while the TM parameters are kept constant. Empirically, this results in better convergence of the training. If $\mathcal{L}_{\text{disc}}<\ln(4)$ we perform one step of classifier training per one step of TM training although this ratio may be tuned to better suit the setup  (e.g., its optimum depends on the learning rate ratio for the discriminator and the TM).

\paragraph{Consistency loss.}
A further regularisation of the method is provided with a so-called \textit{consistency} loss $\mathcal{L}_{cons}$, described in \cite{TURBO,YGAN}. The idea is that the latent representation of $\boldsymbol{\hat{x}}$, which can be obtained by passing it through the same encoder network  $\boldsymbol{\hat{z}}=e_\phi(\boldsymbol{\hat{x}}, \hat{m})$ as shown in Fig.\,\ref{fig:losses}, should be equal to latent representation $\boldsymbol{z}$ from which $\boldsymbol{\hat{x}}$ was created. This can be enforced with the MSE loss between these latent representations.

\begin{equation}
    \mathcal{L}_{\text{cons}}=\mathbb{E}_{\boldsymbol{z}\sim p(\boldsymbol{Z}), \hat{m}\sim p(\hat{M})}||\boldsymbol{z}-\boldsymbol{\hat{z}}||^2=\mathbb{E}_{\boldsymbol{z}\sim p(\boldsymbol{Z}), \hat{m}\sim p(\hat{M})}||\boldsymbol{z}-e_\phi(d_\theta(\boldsymbol{z}, \hat{m}), \hat{m})||^2,
\end{equation}
where we provide $\hat{m}$ by shuffling the mass batches $\{\hat{m}\}=\sigma_{\text{perm}}(\{m\})$ and $z$ by encoding original event-mass pairs $z=e_\phi(\boldsymbol{x}, m)\sim p_\phi(\boldsymbol{Z})$.

Its first advantage is that if both reconstruction and consistency losses achieve zero simultaneously, the transport can be inverted as 
\begin{equation}
\begin{split}
    & f_{\phi,\theta}(f_{\phi,\theta}(\boldsymbol{x}, m, \hat{m}), \hat{m}, m) \\
    & = d_\theta(e_\phi(d_\theta(e_\phi(\boldsymbol{x}, m),  \hat{m}), \hat{m}), m)  \\
    & = d_\theta(e_\phi(\boldsymbol{\hat{x}}, \hat{m}), m) = d(\boldsymbol{\hat{z}}, m)   \stackrel{\mathcal{L}_{cons}=0}{=} d(\boldsymbol{z}, m) = \hat{x} \stackrel{\mathcal{L}_{rec}=0}{=} x,\\
\end{split}
\label{eq:invert1}
\end{equation}
meaning the transport function is round-trip reversible \(f_{\phi,\theta}(\cdot, \hat{m}, m) = f_{\phi,\theta}^{-1}(\cdot, m, \hat{m})\).
Although the consistency loss is not the only way to enforce round-trip reversibility,\footnote{Round-trip reversibility may also be enforced via an explicit loss term, \(\lVert f_{\phi,\theta}(f_{\phi,\theta}(\boldsymbol{x}, m, \hat{m}), \hat{m}, m) - \boldsymbol{x} \rVert\).} the reconstruction and adversarial losses alone do not guarantee round-trip reversibility, as demonstrated by a counterexample in App.\;\ref{app:conter_reverce}.
Furthermore, round-trip reversibility is a sufficient condition for the transport function to be invertible (i.e., bijective) for any fixed \(m\) and \(\hat{m}\), but it is not a necessary condition, as is also shown in App.\;\ref{app:conter_reverce}.
In App.\;\ref{app:proof}, we then prove that when \(\mathcal{L}_{\text{rec}} = 0\) and \(\mathcal{L}_{\text{cons}} = 0\), under our specific decomposition of the transport model (TM), the transport function becomes transitive; that is, \(f_{\phi,\theta}(f_{\phi,\theta}(\boldsymbol{x}, m, \tilde{m}), \tilde{m}, \hat{m}) = f_{\phi,\theta}(\boldsymbol{x}, m, \hat{m})\) for any intermediate \(\tilde{m}\).

The second advantage relies on the adversarial discriminator loss to achieve values close to maximum while minimising consistency and reconstruction losses. 
If $\mathcal{L}_{\text{disc}}=\ln(4)$ with a sufficiently good classifier, we can assume approximate equality between generated and data joint probability distributions $\hat{p}_{\phi,\theta}(\hat{X}, \hat{M})\approx p_{\text{data}}(X, M)$. 
In App.\;\ref{app:proof}, we prove that the equivalence between these distributions, the round-trip reversibility and transitivity lead to the independence of $\hat{x}$ and $m$.
Consequently, $\boldsymbol{\hat{z}}=e_\phi(\boldsymbol{\hat{x}}, \hat{m}) \perp m$\footnote{In our notation the $\perp$ sign denotes statistical independence between two variables.} as any function on variables independent on $m$ returns a variable independent of $m$, and for a zero consistency loss $\boldsymbol{z}=\boldsymbol{\hat{z}} \perp m$. 
Thus, by minimizing \(\mathcal{L}_{\text{rec}}\) and \(\mathcal{L}_{\text{cons}}\) while simultaneously maximizing \(\mathcal{L}_{\text{disc}}\), we approach mass decorrelation in the latent variables \(\boldsymbol{z}\), meaning that the latent representation \(\boldsymbol{z}\) will have approximately the same distribution across any mass range within the training region.
However, in our case, no prior is imposed on the latent distribution, unlike in Variational Autoencoders (VAEs) or Normalizing Flows, where a prior is explicitly defined.
As a result, the model is free to learn any form of latent space distribution.

Although the latent feature mass decorelation is a main conceptual advantage of consistency loss, we show empirically that it also helps to improve the quality of the transport. 
Results shown in App.\,\ref{app:consistency_benefits} confirm that including consistency loss in optimisation provides both of these benefits.

Additionally, for computing the consistency loss $\mathcal{L}_{\text{cons}}$, we use the masses $\hat{m}$ not only from the SB but also from the SR.
In this way, the round-trip reversibility of the transport is also ensured in the region between the two sidebands, connecting all three regions and achieving high-quality interpolation.

\paragraph{Full loss.} Finally, for training of the TM, we combine all losses with their corresponding weights $w_{\text{rec}},\;w_{\text{disc}}$ and $w_{\text{cons}}$, namely
\begin{equation}    \mathcal{L}_{TM}=w_{\text{rec}}\mathcal{L}_{\text{rec}}-w_{\text{disc}}\mathcal{L}_{\text{disc}}+w_{\text{cons}}\mathcal{L}_{\text{cons}}.
\end{equation}

For interpretability, we prioritise the transport to be fixed to identity for $\hat{m}=m$; thus, we assign the highest weight to the reconstruction loss, namely $w_{\text{rec}}=1$. $\mathcal{L}_{\text{rec}}$ and $\mathcal{L}_{\text{cons}}$ are of the same order of magnitude, as both are based on MSE, so we assign $w_{\text{cons}}=0.1$. $\mathcal{L}_{\text{disc}}$ has a different behaviour; thus an appropriate value for $w_{\text{disc}}$ is determined empirically.

\subsection{Architecture}
\label{subsec:arch}

\begin{figure}
    \centering
    \includegraphics[width=1\linewidth]{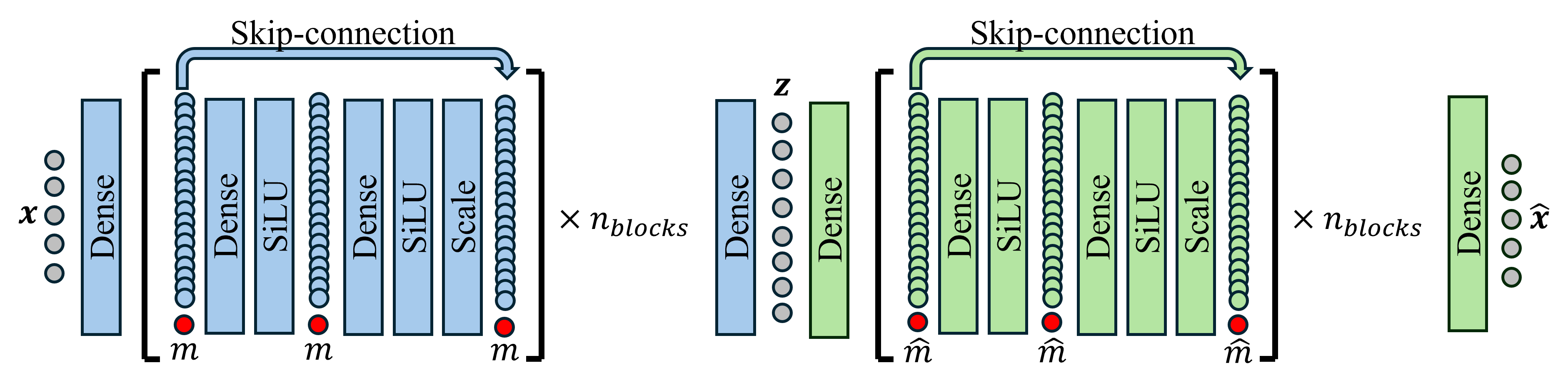}
    \caption{Architecture of the encoder (light-blue) and decoder (light-green) networks in TRANSIT.}
    \label{fig:arch}
\end{figure}

In order to achieve maximal training time efficiency, the network architecture has to be adapted to match the task. 
We are interested in transporting the events between two distributions that are relatively close to each other, and we want mass-decorrelated variables to remain unchanged, thus, we use an MLP with a residual architecture shown in Fig. \ref{fig:arch}. 
The skip-connections combine the input of the residual block $j$ with a scaled output of a residual block $j$ as $\boldsymbol{y}_j=\boldsymbol{x}_{\text{inp},j}+\boldsymbol{\alpha}_j\odot \boldsymbol{f}_{\text{block},j}(\boldsymbol{x}_{\text{inp},j}, m)$, such that the identity transformation is easily learnable by setting learnable parameters $\boldsymbol{\alpha}_j$ equal to 0. This way $\boldsymbol{\alpha}_j \odot\boldsymbol{f}_{\text{block},j}(\boldsymbol{x}_{\text{inp}}, m)$ represents a small mass-conditional correction to the input. Additionally, the latent space vector $\boldsymbol{z}$ has higher dimensionality than input $\boldsymbol{x}$, thus ensuring that the network has no informational bottlenecks, unlike usual auto-encoders.

To make the transport curves smooth, it suffices to use Sigmoid Linear Unit instead of the usual Rectified Linear Unit (ReLU) as we observe empirically\footnote{One can also achieve a smoothing effect by adding a loss based on the average second derivative of the transport curve, however, this requires more computation.}. 
The adversarial discriminator is a simple conditional MLP with ReLU activations.

Conditioning is applied in every dense layer of decoder, encoder and discriminator by appending $m$ or $\hat{m}$ to the input of each linear layer.

\subsection{Anomaly detection strategies}

Optimising the speed of the template generation algorithm is beneficial as long as it remains more resource-intensive than the rest of the anomaly detection pipeline.
Therefore, we use a CWoLa classifier based on Boosted Decision Trees, which proved to be both fast and performant in \cite{Finke:2023ltw,Freytsis:2023cjr}.
We use the same hyperparameters as \cite{RADOT} for a straightforward result comparison.
The same TRANSIT model can be used in two different anomaly detection approaches.

First, to create a template, one can sample events $(\boldsymbol{x}, m)$ from the SB and transport them into the SR by decoding them with masses $\hat{m}$ sampled from the SR. To produce the template, we bootstrap-resample four times as many mass points as there are data points in the SR in total, as recommended in \cite{CURTAINSF4F}.
We then train CWoLa, assuming the created template is signal-poor (label 0) and the data from the signal region is relatively signal-rich (label 1).
This is the default approach and will further be referred to as TRANSIT as well.

In a second approach, we transform both the SB and the SR into latent space.
As the latent space variables are uncorrelated with mass (for the background distribution), classical SB-versus-SR CWoLa training can be used.
This classifies the latent representation of SB data as the signal-poor template (label 0) versus the latent representation of the SR data (label 1).
This method is referred to as latent TRANSIT (LaTRANSIT).\footnote{In analogy with the LaCATHODE method \cite{LaCATHODE}.}

%% file: tex/results.tex
\label{sec:results}

\subsection{Template quality}
\label{subsec:quality}
\begin{figure}[h]
    \centering
    \includegraphics[width=1\linewidth]{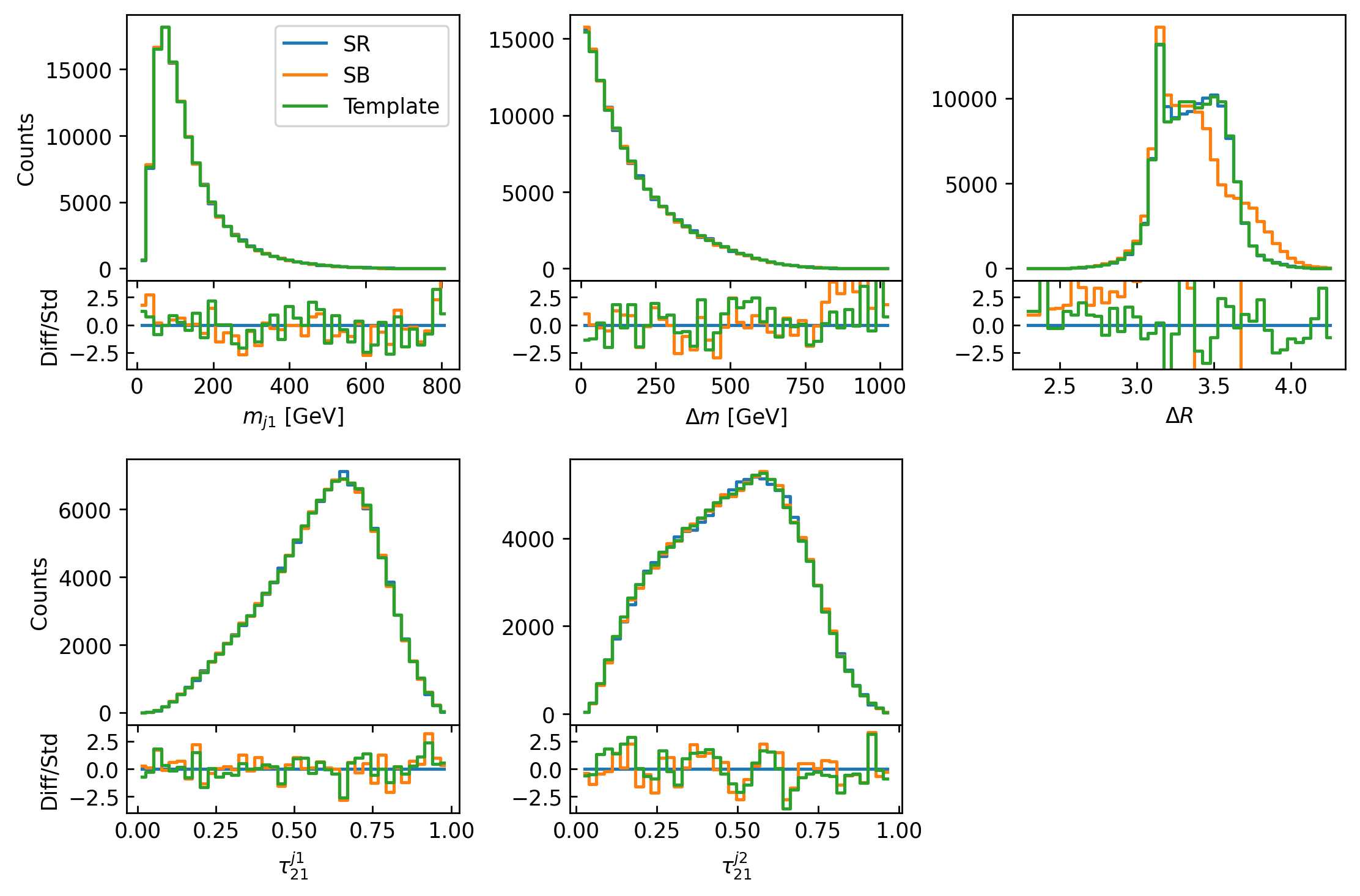}
    \caption{Distributions of five observables for the SR, SB, and a TRANSIT template created by transporting SB events into SR masses. Pull plots illustrate the difference between the SR distribution and the other distributions, expressed in units of the Poisson standard deviation for each bin.}
    \label{fig:marginals}
\end{figure}

After generating the template in the SR, we can compare it to the actual background data in the SR.
Fig.\,\ref{fig:marginals} shows that the template distribution created by transporting SB events into SR mass matches the template distribution closely (within 2.5 $\sigma$ or less in the majority of the bins). This holds even for the $\Delta R$ observable, where SB and SR strongly differ. 
It is noteworthy that our method reconstructs the sharp peak at $\Delta R\approx3.2$, which is a challenging task for simple interpolators like RAD-OT\,\cite{RADOT}. 
Additionally, we verify that TRANSIT produces correct marginal distributions and pairwise correlations between variables when transporting from the lower to the higher sideband and vice versa, as shown in App.\,\ref{app:SBtoSB}.

\begin{figure}[h]
    \centering
    \includegraphics[width=0.5\linewidth]{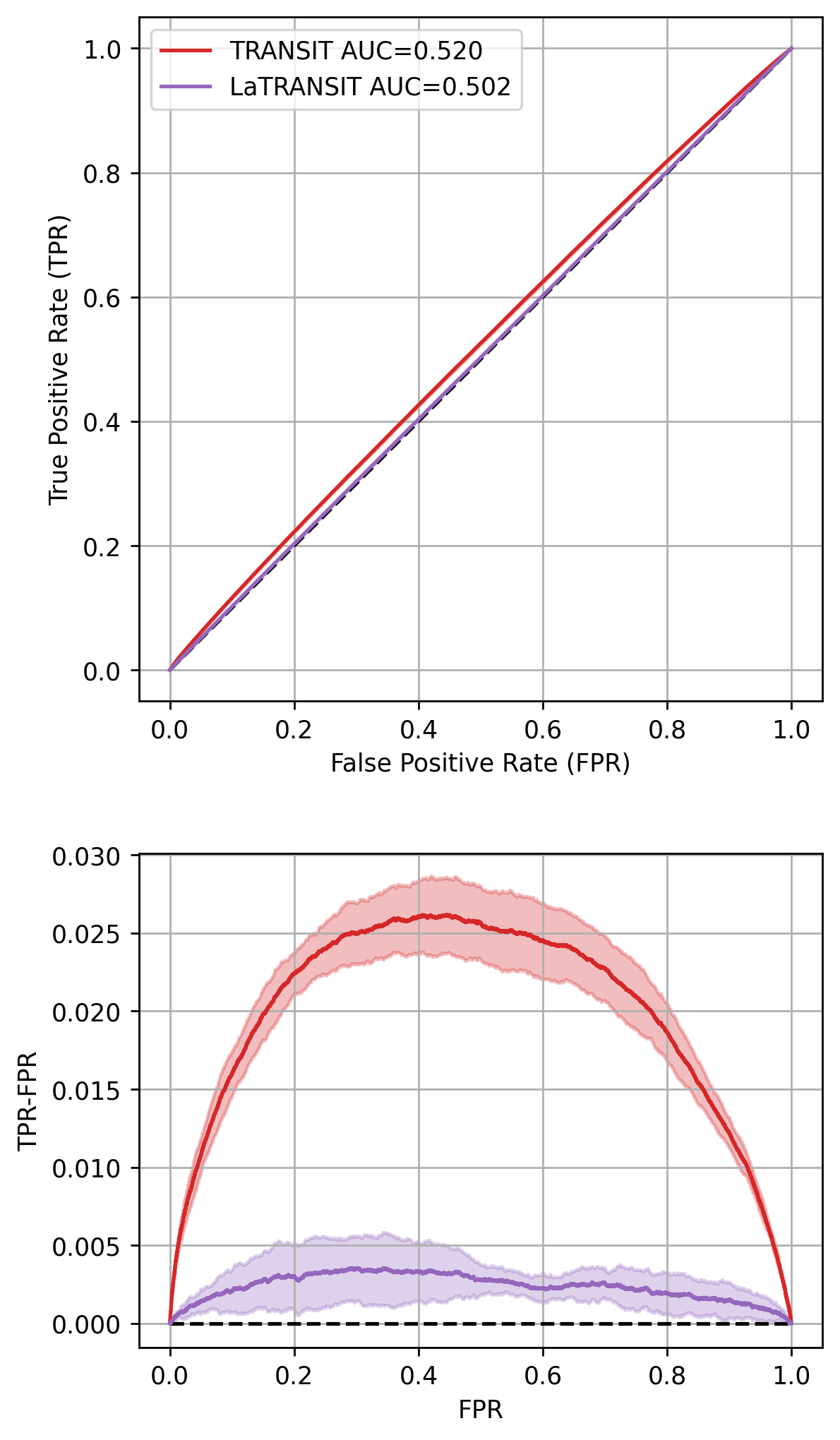}
    \caption{
    ROC curves for a BDT trained to discriminate TRANSIT templates from background SR data and for a BDT trained to discriminate SB latent representations from background SR latent representations in LaTRANSIT. 
    Solid lines and filled regions represent the average and the standard deviation range across 6 TRANSIT network trainings with different initialisation seeds. No signal was added in these runs.}
    \label{fig:closure}
\end{figure}

To assess the overall quantitative similarity between the distribution of true data and that of transported events, we employ a classifier test.
Namely, we train a BDT classifier to discriminate between the template and background data in the SR.
Fig.\,\ref{fig:closure} shows that the two distributions are indeed close, as the receiver operating characteristic (ROC) curve of this classifier is close to the ROC curve of a random classifier.
The area under the ROC curve (AUC) for TRANSIT is only 0.520, which is similar to the values quoted for other methods such as CURTAINSF4F and RAD-OT, as given in \cite{RADOT}.
We emphasise that smooth nonlinear interpolation is not a well-defined problem and, thus, we expect all interpolation methods to match the signal region distribution with limited precision.
We can also look at the adversarial MLP discriminator that is trained as part of the TRANSIT model.
At the end of training, the discriminator reaches a plateau where the loss stochastically fluctuates around $\mathcal{L}_{disc}=\ln(4)$, meaning the discriminator cannot distinguish true events from the transported ones.
In this state, the scores of this classifier for the transported events should be close to 0.5.
This is indeed the case as observed in Fig.\,\ref{fig:traj} in App.\,\ref{app:trajectories} along each of the transport trajectories.

The smoothness and non-linearity of the transport trajectories can also be visually inspected in Fig.\,\ref{fig:traj} of App.\,\ref{app:trajectories}.

\subsection{Decorrelation}
\label{subsec:decor}
As discussed in Section \ref{sec:method}, the TRANSIT training scheme leads to independence between latent variables $\boldsymbol{z}$ and the mass $m$, meaning, $p(\boldsymbol{z}, m)=p(\boldsymbol{z})p(m)$. 

On one hand, independence implies that in any selected mass range, the distribution $p(\boldsymbol{z})$ for the events in this mass range should remain the same.
We apply a classifier test by training a BDT to compare the distributions of $p(\boldsymbol{z})$ in SB and SR. As demonstrated in Fig.\,\ref{fig:closure}, the classifier differs narrowly from a random classifier, thus validating the independence of $p(\boldsymbol{z})$ on the mass region. 

\begin{figure}[h]
    \centering
    \includegraphics[width=0.8\linewidth]{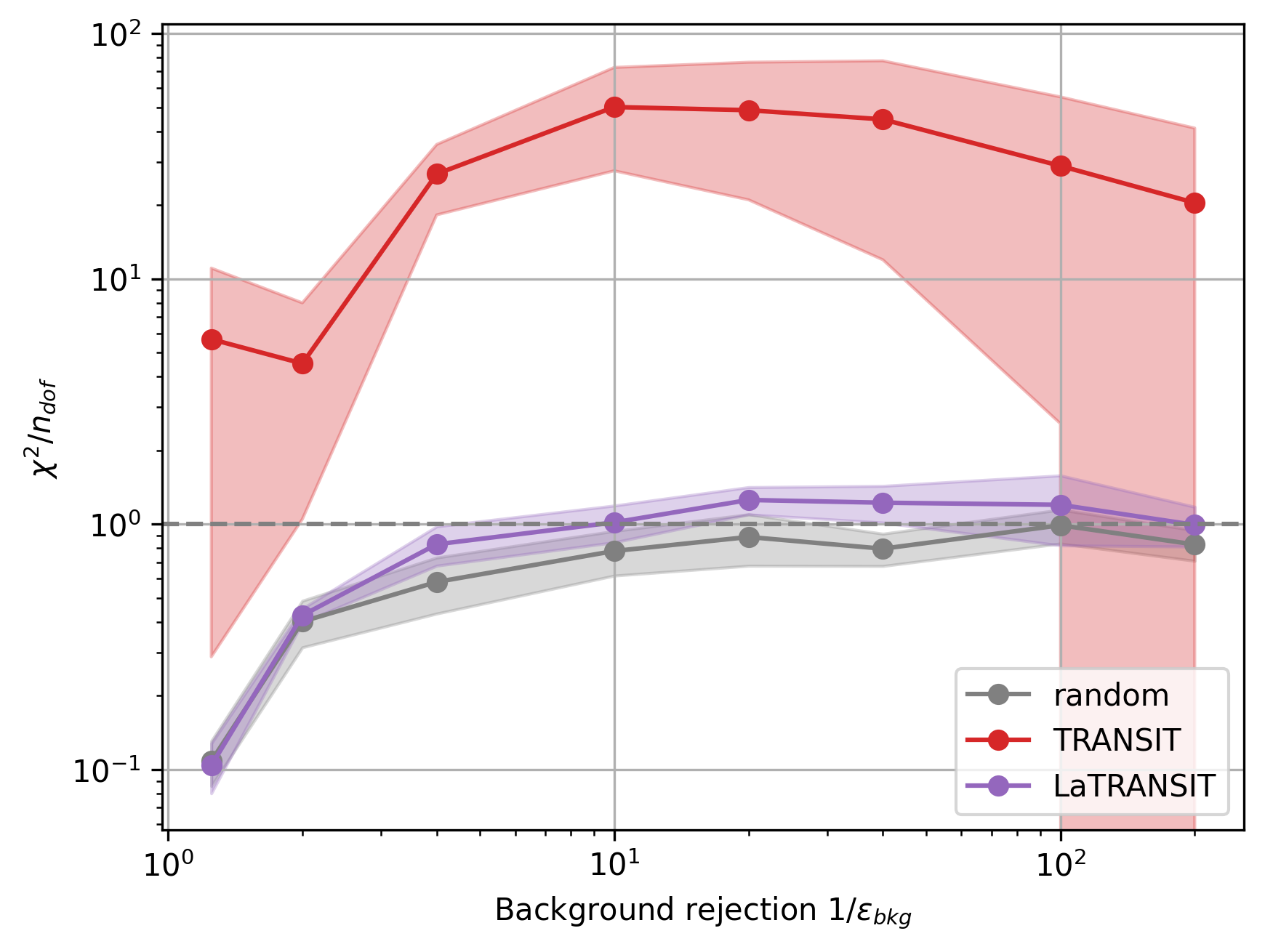}
    \caption{The $\chi^2/n_{d.o.f.}$ discrepancy between the normalised mass distribution of all background events in the region SB$\cup$SR and the distribution of a selection of these events, made based on the TRANSIT anomaly score, LaTRANSIT anomaly score, or a random selection. The calculation of $\chi^2/n_{d.o.f.}$ is done with a histogram of 40 equal width bins in $[3.0, 4.6]$ TeV range. Solid lines and filled regions represent the average and the standard deviation for 6 runs with different random seeds and no signal contamination. No signal was added in these runs. }
    \label{fig:sculpting_chi}
\end{figure}

On the other hand, for any region in $\boldsymbol{z}$-space, the events in it should have the same distribution of mass $m$ if $\boldsymbol{z}$ and $m$ are independent. 
A score of a classifier trained to distinguish between SB and SR in the latent space only depends on variable $\boldsymbol{z}$, and thus, selecting the event with the score larger than some threshold should not change the mass distribution significantly. 
Fig.\,\ref{fig:sculpting_chi} shows the $\chi^2/n_{d.o.f.}$ difference between the original mass spectrum $p_M(m)$ and the spectrum after a classifier cut.
For a random classifier, this difference increases until it reaches $\chi^2/n_{d.o.f.}=1$, which is the expected discrepancy for two independent samples of the same distribution.
Since the TRANSIT method uses variables with significant mass correlation (e.g., $\Delta R$), the classifier score is also mass-dependent, and a cut on this score induces strong background sculpting even for small rejections.
In contrast, a LaTRANSIT cut has approximately the same effect on the distribution as a random cut, proving the independence of this classifier score from the event mass.
Appendix \ref{app:bkg_sculpting} also visually demonstrates the presence of the background sculpting effect for the TRANSIT method and its absence in the LaTRANSIT method.

\subsection{Anomaly detection}

To assess the anomaly detection performance of the proposed method and compare it to benchmarks in the literature, we perform the analysis with an injection of $N_{\text{sig}}$ events into our background sample.  
Most of the signal events land in the selected SR.  
As discussed before, we choose RAD-OT as one of the fastest methods and CURTAINSF4F as one of the highest-quality template generation methods for comparison.\footnote{Moreover, the authors of these methods provide sufficient details needed to ensure a fair comparison.}  

\begin{figure}[h]
    \centering
    \includegraphics[width=0.49\linewidth]{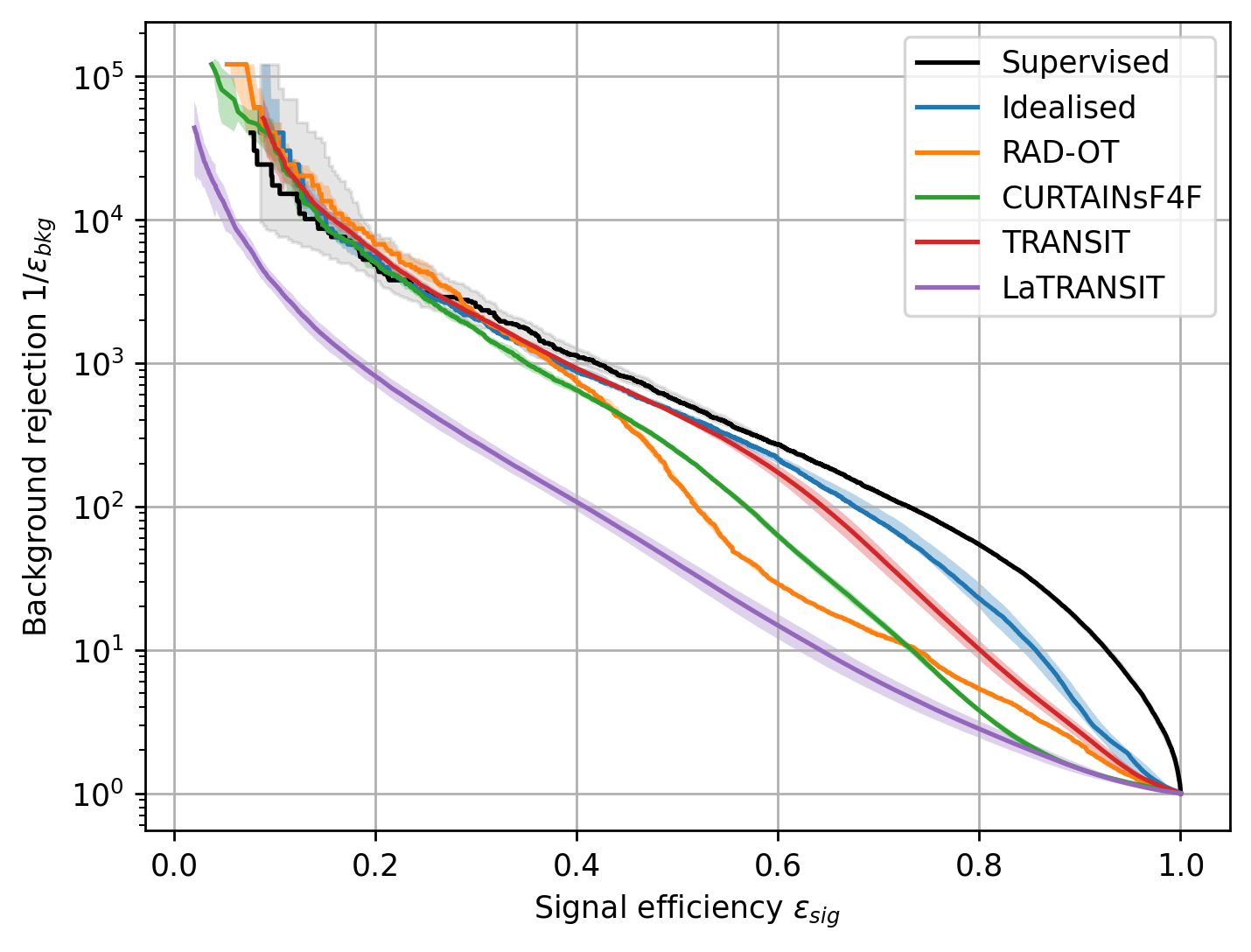}
    \includegraphics[width=0.49\linewidth]{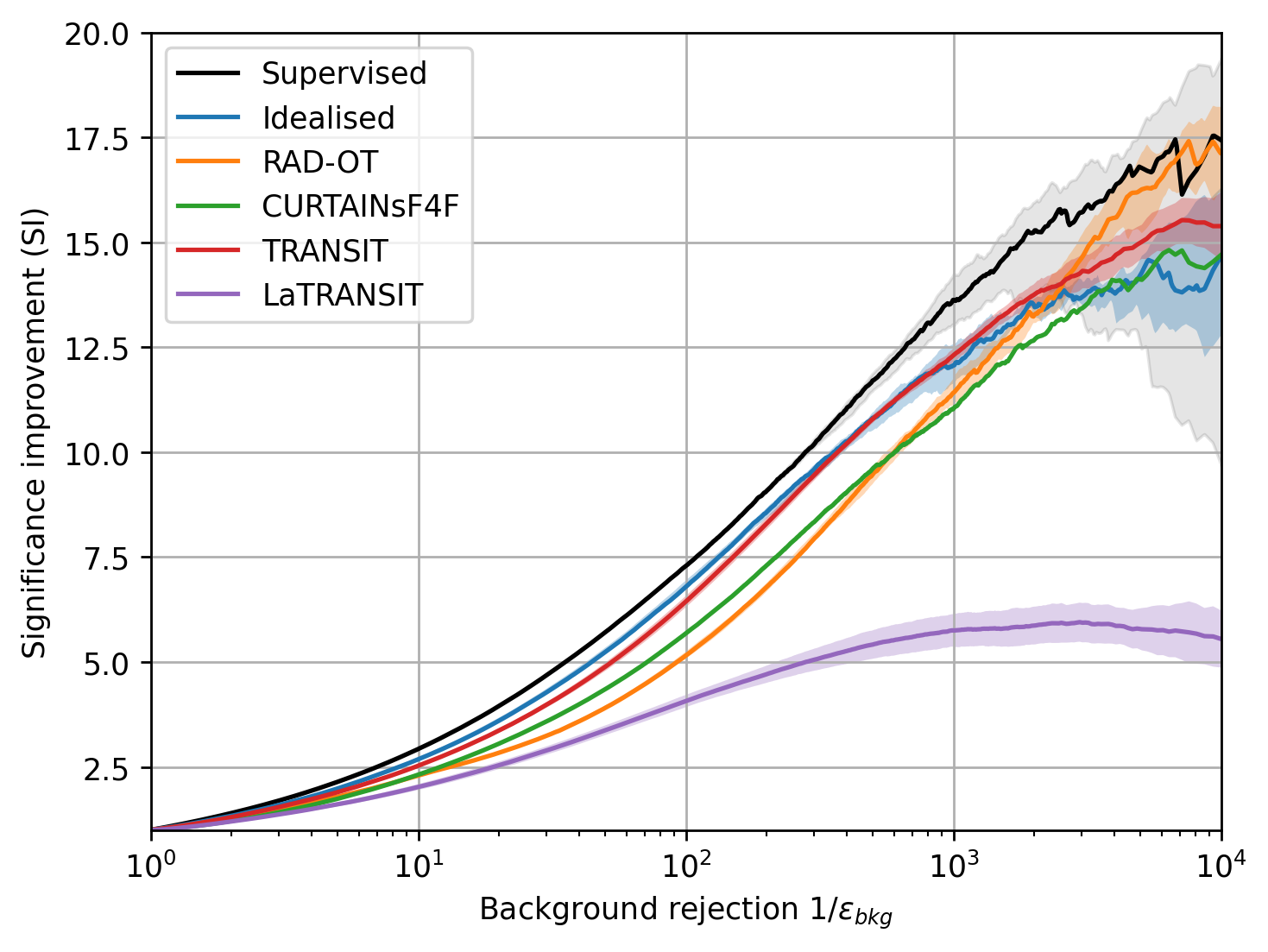}
    \caption{Background rejection as a function of signal efficiency (right) and significance improvement as a function of background rejection (left) compared for various methods.  
    The results are produced by injecting 3000 $Z'$ signal events.  
    Solid lines and filled regions represent the average and the standard deviation range for 30 TRANSIT network trainings with different initialisation seeds.  
    For supervised, idealised, RAD-OT, and CURTAINsF4F, the average and the standard deviation are taken by retraining the CWoLa classifier 5 times with various seeds. }
    \label{fig:rej_SI}
\end{figure}

\begin{figure}[h]
    \centering
    \includegraphics[width=0.49\linewidth]{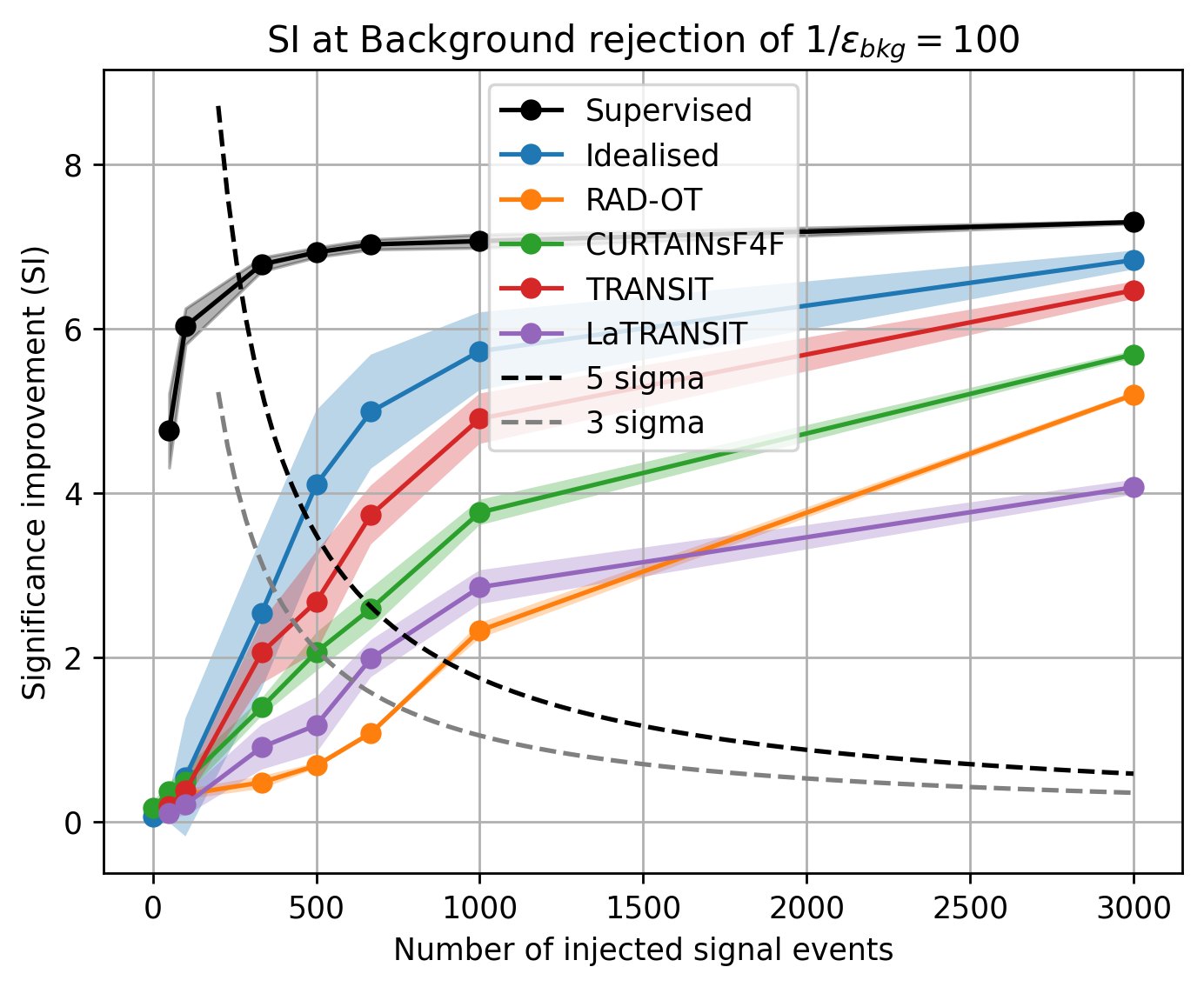}
    \includegraphics[width=0.49\linewidth]{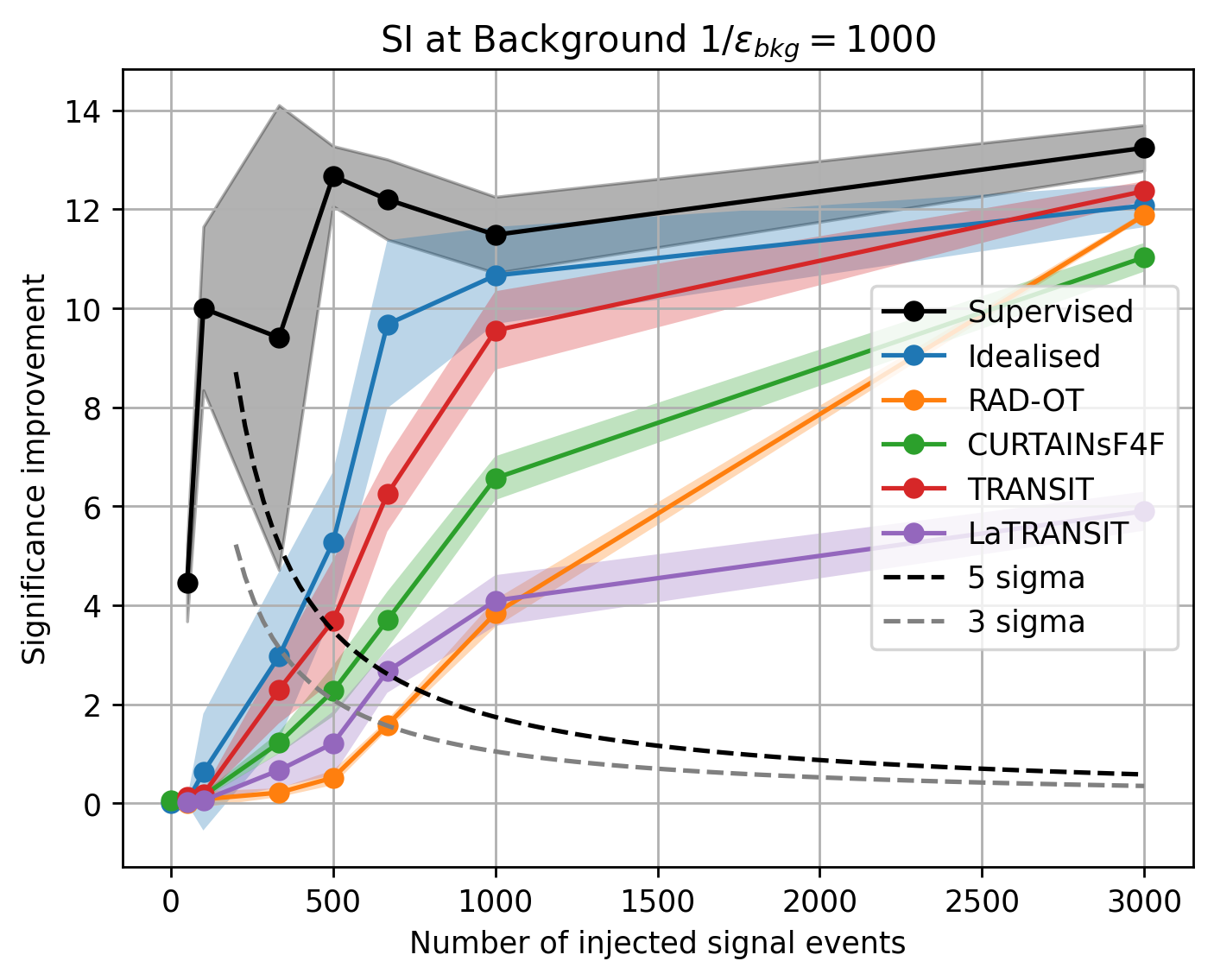}
    \caption{Significance improvement as a function of the number of injected signal samples using a background rejection value of 100 (right) and 1000 (left), compared for various methods.  
    Solid lines and filled regions represent the average and the standard deviation range for 6 TRANSIT network trainings with different initialisation seeds.  
    For supervised, idealised, RAD-OT, and CURTAINsF4F, the average and the standard deviation are taken by retraining the CWoLa classifier 5 times with various seeds.}
    \label{fig:SI_dope}
\end{figure}

Additionally, we use two upper bounds on the performance of our method.  
First, we train a classifier to distinguish a pure background from a pure signal in a \textit{supervised} manner.  
Since we use exact, noise-free labels for this method, its performance is expected to be higher than that of any semi-supervised method.  
The \textit{idealised} CWoLa variant represents a case of perfect template generation in a CWoLa-like search.  
In this method, we use half of our background events as the template and the other half with the injected signal as our data.  
This way, the background in both datasets is sampled from the same distribution.  
This version is expected to perform worse than the \textit{supervised} method but better than any semi-supervised approach that uses the same statistics in the SR data and template.
As proven in \cite{CURTAINSF4F}, semi-supervised methods gain improved performance at high rejection rates by sampling more template events in the SR region than there are SR data events — a technique referred to as "oversampling."
The template generation methods presented here use a fourfold oversampling strategy. 
As shown in Fig.\,\ref{fig:rej_SI}, this allows them to achieve slightly higher SI at high rejections than the \textit{idealised} method, whose statistics are limited by the dataset size.
The relatively small difference between the \textit{supervised} and \textit{idealised} methods in Figs.\,\ref{fig:rej_SI}, and \ref{fig:SI_dope} indicates the robustness of the CWoLa classifier to noisy labels.  
However, neither of these two methods can be used in practice, as the labels are unknown.  

First, we inject $N_{\text{sig}} = 3000$ signal events.  
Fig.\,\ref{fig:rej_SI} shows the relation between background rejection and signal efficiency and the relation between significance improvement and background rejection (higher curves indicate better performance) for all methods.\footnote{The results for non-TRANSIT methods are taken from Ref.\,\cite{RADOT} with permission of the authors.}  
One can clearly see that for low signal efficiency, the performance of all ML methods except LaTRANSIT saturates and reaches that of the \textit{supervised} and \textit{idealised} bounds.  
However, in an analysis, we aim to preserve most of the signal while rejecting a substantial portion of the background, so we are interested in signal efficiencies higher than 0.4 and, consequently, background rejections lower than $10^3$.  
In this region, one can see that the TRANSIT method outperforms both RAD-OT and CURTAINsF4F and, overall, has performance close to the \textit{idealised} case.  
At the same time, LaTRANSIT exhibits lower anomaly detection performance than the rest of the methods.  
Since the invariant mass $m_{jj}$ is a defining feature of a resonance, it has high discriminative power in a signal-versus-background classifier.  
Thus, any strongly mass-correlated variable, such as $\Delta R$, also enhances classifier performance.  
This explains why LaTRANSIT, where we use only mass-decorrelated observables, inevitably has lower performance than the other methods.  

However, a method is even more valuable if it retains sensitivity for a small number of signal events.  
We present SI as a function of $N_{\text{sig}}$ in Fig.\,\ref{fig:SI_dope} for a classifier cut with background rejection of 100 and 1000 for the described methods.\footnote{The results for non-TRANSIT methods are taken from Ref.\,\cite{RADOT} with permission of the authors.} 
Assuming we can perfectly estimate the background count in SR, a simple counting experiment in this region would provide a significance of $Z=N_{\text{sig,SG}}/\sqrt{N_{\text{bkg,SG}}}$.This means that to detect evidence of a signal, $Z_{\text{evid}}=3\sigma$, we need a significance improvement of $SI\ge Z_{\text{evid}}\sqrt{N_{\text{bkg,SG}}}/N_{\text{sig,SG}}$, which is shown in Fig.\,\ref{fig:SI_dope} as a gray dashed line.  
The black dashed line analogously shows the threshold at which a discovery could be claimed with $Z_{\text{disc}} = 5\sigma$.  

We observe that the generation of TRANSIT templates yields a higher SI than both RAD-OT and CURTAINSF4F  across the range of signal contaminations where at least one of the methods achieves $SI > 1$.  
The TRANSIT curve intersects the evidence and discovery thresholds at a lower number of $N_{\text{sig}}$, thereby having greater discovery potential.  
It also comes close to the performance of the \textit{idealised} template generator, generally exhibiting SI values that lay in the $2.5$ s.d. range of \textit{idealised} classifier.  

Despite LaTRANSIT having the lowest SI performance for a high $N_{\text{sig}}$ in a given background rejection, it actually outperforms RAD-OT in the region of interest, namely below the $5\sigma$ threshold.
More importantly, comparing the two subplots in Fig.\;\ref{fig:rej_SI} shows that higher background rejection values lead to better SI performance.
As shown in App.\;\ref{app:bkg_sculpting}, the distribution of the resonant variable after a cut based on a method that relies on mass-correlated observables (e.g., TRANSIT) becomes increasingly sculpted for higher background rejections. This hinders the fitting of a background fit function, which is often assumed to be smoothly falling, thus restricting the application of methods based on mass-correlated observables to low background rejections.
However, the mass-independent variables in LaTRANSIT prevent background sculpting, allowing the use of high background rejections and thus providing similar or better SI performance than RAD-OT, CURTAINS, and TRANSIT when these are limited to low background rejection values.
The exact analysis performance depends on the chosen background fitting and bump-hunting procedures and is outside the scope of this work.

\subsection{Computational efficiency}

\begin{figure}[h]
    \centering
    \includegraphics[width=0.8\linewidth]{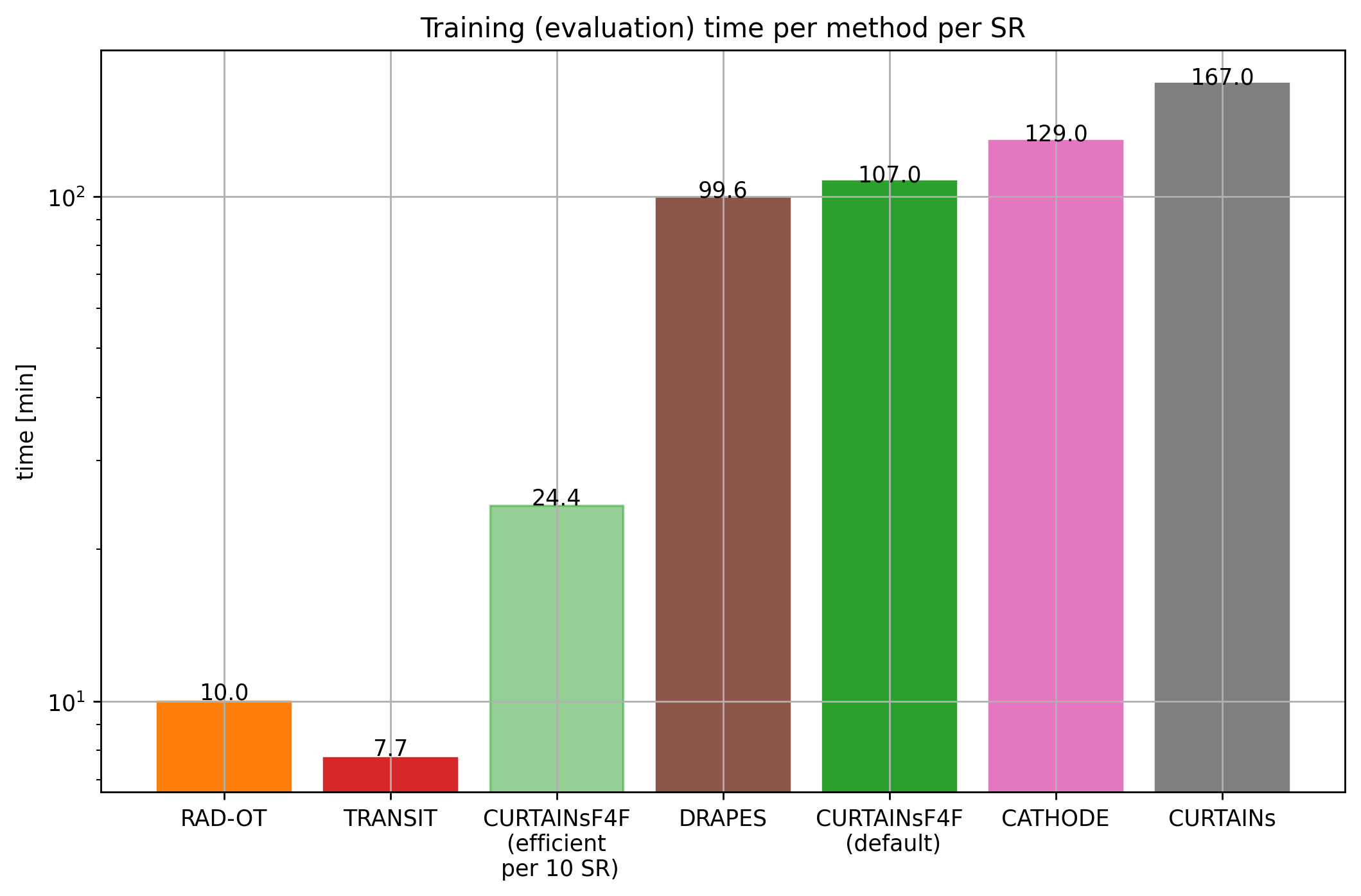}
    \caption{Comparison of the approximate time to train a model (if needed) and generate a template of 50000 events for several template generation methods. RAD-OT uses 1 CPU core, while the rest of the methods utilise 1 GPU and 16 CPU.}
    \label{fig:timing}
\end{figure}

Fig.\,\ref{fig:timing} shows the estimated time required to obtain a template using TRANSIT and other methods for comparison,\footnote{The results for non-TRANSIT methods are taken from references \cite{CURTAINSF4F,RADOT} with permission of the authors.} including the time required to train a model on the sideband data and the time needed to generate 500,000 template events in SR.  
The generation time is usually much shorter than the time required to train an ML model, so we neglect the former for DRAPES, CURTAINsF4F, CURTAINs, and CATHODE.  
RAD-OT \cite{RADOT} does not require training, and the template is computed using one CPU core.  
For all other methods, training and generation were performed using one $\text{NVIDIA}^{\text{\tiny{®}}}$ RTX 3080 GPU and 16 CPU cores for parallel data loading.  
Efficient CURTAINsF4F relies on the additional assumption that the ``base" flow can be trained only once using all the provided data and that only a small ``top" flow needs to be trained for each new signal region.  
Thus, the cost of training the ``base" flow is distributed across all signal regions.  
We used 10 signal region windows, which is a representative order of magnitude for dijet analyses.  

It is evident that TRANSIT achieves more than a tenfold speedup compared to most other ML methods.  
It is significantly faster than the efficient CURTAINsF4F version for a moderate number of SR windows.  
TRANSIT also achieves a template generation time comparable to RAD-OT; however, it utilises more computational resources. 

%% file: tex/conclusion.tex
\label{sec:conclusion}

In this work, we developed TRANSIT, a method for conditional data transport and ``implicit" condition decorrelation based on adversarial neural network training.  
The method was applied to the problem of data-driven generation of background templates for semi-supervised, model-agnostic anomaly searches in high-energy physics and evaluated using the LHCO R\&D dataset.  
Our results show that TRANSIT is capable of smooth and non-linear interpolation of data, creating a high-quality template that closely mimics the background in the signal region.  
When integrated into the CWoLa framework, TRANSIT achieves competitive anomaly detection performance, substantially outperforming non-ML-based methods such as RAD-OT \cite{RADOT} and surpassing prior transport-based deep learning methods such as CURTAINSF4F \cite{CURTAINSF4F}.  
Additionally, TRANSIT requires an order of magnitude less training time than many flow- and diffusion-based models.  

One of the most significant insights from this work is that high-quality template generation for weakly supervised searches can be achieved without resorting to complex flow- or diffusion-based models.  
By simply setting the right optimisation objectives and employing appropriate loss functions, we demonstrated that it is possible to achieve both high performance and computational efficiency.  
Moreover, the strategy of transporting events instead of generating them from scratch, coupled with an architecture specifically designed to streamline this process, has resulted in a remarkably efficient model.  In conclusion, TRANSIT’s simplicity and speed make it a highly scalable solution, capable of handling the computational demands of modern anomaly search analysis pipelines.  

Another key feature of TRANSIT is its ability to make latent space variables independent from the invariant mass condition after a convergent training.  
This enables anomaly searches to be performed in the space of mass-decorrelated variables, which we have referred to as LaTRANSIT.  
Despite lower significance improvements for a given background rejection value, LaTRANSIT has high robustness to mass sculpting, providing a beneficial trade-off in the analysis context.  
Most semi-supervised methods, including TRANSIT, should only be utilised for low background rejection values in order to preserve the shape of the mass spectrum so that it can be fitted in later analysis stages.  
However, methods such as LaCATHODE \cite{LaCATHODE} and LaTRANSIT provide the possibility to set much higher rejection working points, corresponding to greater analysis sensitivity, without suffering from an increased false discovery probability.  

The approach is not limited to low-dimensional tabular data, as the dense networks in the encoder, decoder, and discriminator components could all be replaced with a suitable architecture for a different data representation.  
Future work could explore the use of transformers to conditionally morph particle clouds.  
This could prove to be useful for template generation in anomaly searches with low-level observables or for unfolding tasks.  
An alternative direction is to merge the fast training of TRANSIT with efficient multi-signal-region interpolation in ``efficient" CURTAINSF4F \cite{CURTAINSF4F} and SIGMA \cite{SIGMA} to achieve even greater speedups.

%% file: tex/app.tex
\section{Proof of the efficiency of transport models over Normalising Flows}
\label{app:transport_vs_generation}

A conditional generative model that can usually be described by a function  $g(\boldsymbol{\eta}, m) \rightarrow \boldsymbol{x} \sim p(\boldsymbol{X} | M=m)$
where \( \boldsymbol{\eta} \) is sampled from a zero-mean, unit variance Gaussian and \( m \) is sampled from a known distribution \( p(M) \).  
In many generative models, this function is invertible, for example, in DDIM, Normalising Flows, and Continuous Flow Matching or quasi-invertable, as in VAE, CycleGAN, meaning that we can return to the latent space representation using $g^{-1}(\boldsymbol{x}, m) = \boldsymbol{\eta} \sim \mathcal{N}(\boldsymbol{0}, \boldsymbol{1})$.  
Thus, any of these models or a model with a specially learned inverse function, can be turned into a transport model using the relation $f(\boldsymbol{x}, m, \hat{m})=g(g^{-1}(\boldsymbol{x}, m), \hat{m}) = \boldsymbol{\hat{x}} \sim p(X|M=\hat{m})$
for \( \boldsymbol{x} \sim p(\boldsymbol{X} | M=m) \).  

Consider the space of all possible architectures and methods for creating a high-quality transport model $f(\boldsymbol{x}, m, \hat{m})$. Given hardware and data constraints, each method is assigned a specific training time $t_{transport}$. The methods for creating a transport model by repurposing an invertible generative model form a subset of this space, with times $t_{generative}$ equal to the time needed to train such a generative model along with its inverse.
Therefore, if we were to find a way to create a transport model in the minimum possible time, it would require no more than the time needed for the fastest training of an invertible generative model, namely: $min(t_{transport})\leq min(t_{generative})$.

Thus, we have shown that training a transport model in the optimal case is more efficient than, or at least as efficient as, the optimal training of many popular conditional generative models, including Normalising Flows.
Models that do not have an explicit inverse, such as GANs and DDPMs, have analogues with an inbuilt (pseudo-)inverse, such as CycleGAN and DDIM, which have a similar training cost. Therefore, these generative models are also expected to be less cost-efficient than transport training. 

\section{Proof of independence between $\boldsymbol{\hat{X}}$ and $M$ in optimal TRANSIT network}  
\label{app:proof}

Consider an arbitrary point \( (\boldsymbol{x}_1, m_1) \) and masses \( m_2, m_3 \).  
Let us define the transport function for the model described in Section \ref{sec:method} as  
\begin{equation}
    \boldsymbol{x}_2 = f_{m_1, m_2}(\boldsymbol{x}_1) \overset{\text{def}}{=} f(\boldsymbol{x}_1, m_1, m_2) \overset{\text{def}}{=} d_\theta(e_\phi(\boldsymbol{x}_1, m_1), m_2). 
\end{equation}  
We put $m_1$ and $m_2$ as indices to emphasis that in this appendix we consider $f_{m_1, m_2}(\boldsymbol{x})$ as a function of only vector $\boldsymbol{x}$, and different parameters $m_1$ and $m_2$ denote different functions in particular $f_{m_1, m_2}\neq f_{m_2, m_1}$. 
In the case where \( \mathcal{L}_{\text{rec}} \) and \( \mathcal{L}_{\text{cons}} \) are zero, $f_{m_1, m_2}$ is invertible with the inverse given by the transport from mass $m_2$ to mass $m_1$
\begin{equation}
   f^{-1}_{m_1, m_2}(\boldsymbol{x}) = f_{m_2, m_1}(\boldsymbol{x}),
\end{equation}  
due to Eq.\,\ref{eq:invert1}, and is therefore bijective. Analogously, we can write  
\begin{equation}
    \boldsymbol{x}_3 = f_{m_1, m_3}(\boldsymbol{x}_1) \overset{\text{def}}{=} d_\theta(e_\phi(\boldsymbol{x}_1, m_1), m_3).
    \label{B3}
\end{equation}  
If the consistency loss $\mathcal{L}_{cons}$ is zero, then \( e_\phi(\boldsymbol{x}_2, m_2) = e_\phi(\boldsymbol{x}_1, m_1) \), and we obtain  
\begin{equation}
    f_{m_2, m_3}(\boldsymbol{x}_2) = d_\theta(e_\phi(\boldsymbol{x}_2, m_2), m_3) = d_\theta(e_\phi(\boldsymbol{x}_1, m_1), m_3) = \boldsymbol{x}_3.
    \label{B4}
\end{equation}  
Thus combining Eq.\;\ref{B3} and Eq.\;\ref{B4}, the transport is transitive 
\begin{equation}
    f_{m_2, m_3}(f_{m_1, m_2}(\boldsymbol{x}_1)) = f_{m_1, m_3}(\boldsymbol{x}_1).  
    \label{eq:transitivity}
\end{equation}  
We use an encoder \( e_\phi \) and a decoder \( d_\theta \), both of which consist only of differentiable functions (as shown in Subsection \ref{subsec:arch}).  
Thus, for specific values of \( m_1 \) and \( m_2 \), we can differentiate the transport function with respect to its first argument to obtain the Jacobian  
\begin{equation}
J_{m_1, m_2}(\boldsymbol{x}) \stackrel{\text{def}}{=} \left| \det\left( \frac{\partial f_{m_1, m_2}(\boldsymbol{x})}{\partial \boldsymbol{x}} \right) \right|.
\end{equation}  
Using the chain rule of differentiation on Eq.\,\ref{eq:transitivity}, we obtain a differentiable form of transitivity:
\begin{equation}
    J_{m_1, m_3}(\boldsymbol{x}_1)=J_{m_2, m_3}(f_{m_1, m_2}(\boldsymbol{x}_1))J_{m_1, m_2}(\boldsymbol{x}_1).
    \label{eq:jacobian_transitive}
\end{equation}
Additionally, according to the rule of probability density function transformation, given a p.d.f. of one variable $p_X(x)$ and a transformation function $y=f(y)$, one can express the p.d.f. for $y$ as
\begin{equation}
    p_Y(y)=p_X(x)\left|\frac{dx}{dy}\right|= p_X(f^{-1}(y))\left|\frac{df^{-1}(y)}{dy}\right|.
\end{equation}
In multiple dimensions, this rule is extended for any invertible smooth vector function of a vector variable with the same input and output dimensions so that it holds
\begin{equation}
    p_{\boldsymbol{Y}}(\boldsymbol{y})=p_{\boldsymbol{X}}(\boldsymbol{x})\left|det\left(\frac{\partial \boldsymbol{x}}{\partial \boldsymbol{y}}\right)\right|=p_{\boldsymbol{X}}(f^{-1}(\boldsymbol{y}))J_{f^{-1}(y)}(\boldsymbol{y}).
\end{equation}
The conditional distributions $p_X(\boldsymbol{x}|\boldsymbol{c})$ and $p_Y(\boldsymbol{y}|\boldsymbol{c})$ relate analogically 
\begin{equation}
    p_{\boldsymbol{Y},\boldsymbol{C}}(\boldsymbol{y}|\boldsymbol{c})=p_{\boldsymbol{X},\boldsymbol{C}}(\boldsymbol{x}|\boldsymbol{c})\left|det\left(\frac{\partial \boldsymbol{x}}{\partial \boldsymbol{y}}\right)\right|=p_{\boldsymbol{X},\boldsymbol{C}}(f^{-1}(\boldsymbol{y},\boldsymbol{c}),\boldsymbol{c})J_{f^{-1}(\boldsymbol{y}|\boldsymbol{c})}(\boldsymbol{y}).
\end{equation}
This applies to our function $f_{m_1,m_2}$ to yield
\begin{equation}
\begin{split}
    p_{\boldsymbol{\hat{X}},\hat{M},M}(\boldsymbol{\hat{x}}|\hat{m},m)=p_{\boldsymbol{X},\hat{M},M}(f_{\hat{m},m}(\boldsymbol{\hat{x}})|\hat{m},m)J_{\hat{m},m}(\boldsymbol{\hat{x}}), \\
    p_{\boldsymbol{X},\hat{M},M}(\boldsymbol{x}|\hat{m},m)=p_{\boldsymbol{\hat{X}},\hat{M},M}(f_{m,\hat{m}}(\boldsymbol{x})|\hat{m},m)J_{m,\hat{m}}(\boldsymbol{x}).
\end{split}
\label{eq:p_trafo}
\end{equation}
Additionally, let us recall that $\hat{m}$ is a shuffled version of $m$ and thus is statistically independent of either $m$ or $x$, meaning
\begin{equation}
    p_{\boldsymbol{X},M,\hat{M}}(\boldsymbol{x}|m,\hat{m})=p_{\boldsymbol{X},M}(\boldsymbol{x}|m)\; \forall \hat{m}.
\label{eq:m_hat_indep}
\end{equation}
However, this ensures that $\hat{m}$ is a shuffled version of $m$ and has the same marginal distribution
\begin{equation}
    p_M(k)=p_{\hat{M}}(k).
    \label{eq:m_m_hat_eq}
\end{equation}
Finally, the maximisation of the discriminator $\mathcal{L}_{disc}$ loss up to a value of $\ln(4)$ makes joint distribution for pairs $(x, m)$ and $(\hat{x}, \hat{m})$ same, and thus
\begin{equation}
    p_{\boldsymbol{X},M}(k,l)=p_{\boldsymbol{\hat{X}},\hat{M}}(k,l)\stackrel{\ref{eq:m_m_hat_eq}}{\Rightarrow{}}p_{\boldsymbol{X},M}(k|l)=p_{\boldsymbol{\hat{X}},\hat{M}}(k|l).
    \label{eq:hat_eq}
\end{equation}
Consequently, we can summaries that
\begin{equation}
\begin{split}
    p_{\boldsymbol{\hat{X}},\hat{M},M}(\boldsymbol{a}| b, c) \\
    \stackrel{\ref{eq:p_trafo}}{=} p_{\boldsymbol{X},\hat{M},M}(f_{b, c}(\boldsymbol{a})| b, c)J_{b, c}(\boldsymbol{a})\\
    \stackrel{\ref{eq:m_hat_indep}}{=} p_{\boldsymbol{X},M}(f_{b, c}(\boldsymbol{a})| c)J_{b, c}(\boldsymbol{a}) \\
    \stackrel{\ref{eq:hat_eq}}{=} p_{\boldsymbol{\hat{X}},\hat{M}}(f_{b, c}(\boldsymbol{a})| c)J_{b, c}(\boldsymbol{a}) \\
    \stackrel{\text{stat.}}{=} \int^{q=\max(M)}_{q=\min(M)} p_{\boldsymbol{\hat{X}},\hat{M}, M}(f_{b, c}(\boldsymbol{a})| c, q)p_M(q)J_{b, c}(\boldsymbol{a})  dq\\
    \stackrel{\ref{eq:p_trafo}}{=} \int^{q=\max(M)}_{q=\min(M)} p_{\boldsymbol{X},\hat{M}, M}(f_{c, q}(f_{b, c}(\boldsymbol{a}))|c, q) J_{c, q}(f_{b, c}(\boldsymbol{a})) J_{b, c}(\boldsymbol{a}) p_M(q) dq \\
    \stackrel{\ref{eq:m_hat_indep}}{=} \int^{q=\max(M)}_{q=\min(M)} p_{\boldsymbol{X}, M}(f_{c, q}(f_{b, c}(\boldsymbol{a}))| q) J_{c, q}(f_{b, c}(\boldsymbol{a})) J_{b, c}(\boldsymbol{a}) p_M(q) dq \\
    \stackrel{\ref{eq:transitivity}}{=} \int^{q=\max(M)}_{q=\min(M)} p_{\boldsymbol{X}, M}(f_{b, q}(\boldsymbol{a})|q) J_{c, q}(f_{b, c}(\boldsymbol{a})) J_{b, c}(\boldsymbol{a}) p_M(q) dq\\
    \stackrel{\ref{eq:jacobian_transitive}}{=} \int^{q=\max(M)}_{q=\min(M)} p_{\boldsymbol{X}, M}(f_{b, q}(\boldsymbol{a})| q) J_{b, q}(\boldsymbol{a}) p_M(q) dq\\
    \stackrel{\ref{eq:m_hat_indep}}{=} \int^{q=\max(M)}_{q=\min(M)} p_{\boldsymbol{X}, \hat{M}, M}(f_{b, q}(\boldsymbol{a})|b, q) J_{b, q}(\boldsymbol{a}) p_M(q) dq\\
    \stackrel{\ref{eq:p_trafo}}{=} \int^{q=\max(M)}_{q=\min(M)} p_{\boldsymbol{\hat{X}}, \hat{M}, M}(\boldsymbol{a}|b, q) p_M(q) dq\\
    \stackrel{stat.}{=} p_{\boldsymbol{\hat{X}},\hat{M}}(\boldsymbol{a}| b)\\
\end{split}
\label{eq:loong}
\end{equation}
Finally
\begin{equation}
\begin{split}
    p_{\boldsymbol{\hat{X}},\hat{M},M}(\boldsymbol{a}| b, c) = p_{\boldsymbol{\hat{X}},\hat{M}}(\boldsymbol{a}| b) \\
    \Rightarrow{} \int^{c=\max(M)}_{c=\min(M)} p_{\boldsymbol{\hat{X}},\hat{M},M}(\boldsymbol{a}| b, c)p_M(b)db = \int^{c=\max(M)}_{c=\min(M)} p_{\boldsymbol{\hat{X}},\hat{M}}(\boldsymbol{a}| c)p_M(b)db \\
    \Rightarrow{} p_{\boldsymbol{\hat{X}},M}(\boldsymbol{a}|c)=p_{\boldsymbol{\hat{X}}}(\boldsymbol{a})\\
    \Rightarrow{} \boldsymbol{\hat{X}} \perp M.
\end{split}
\end{equation}

In case, $\mathcal{L}_{rec}$, $\mathcal{L}_{cons}$ are not zero and $\mathcal{L}_{trans}$ do not reach $\ln(4)$, we only expect to achieve an approximate independence of $\boldsymbol{\hat{X}}$ and $M$ meaning that the remaining dependence is weak.

\section{Sideband to sideband transport}
\label{app:SBtoSB}

\begin{figure}[H]
    \centering
    \includegraphics[width=\linewidth]{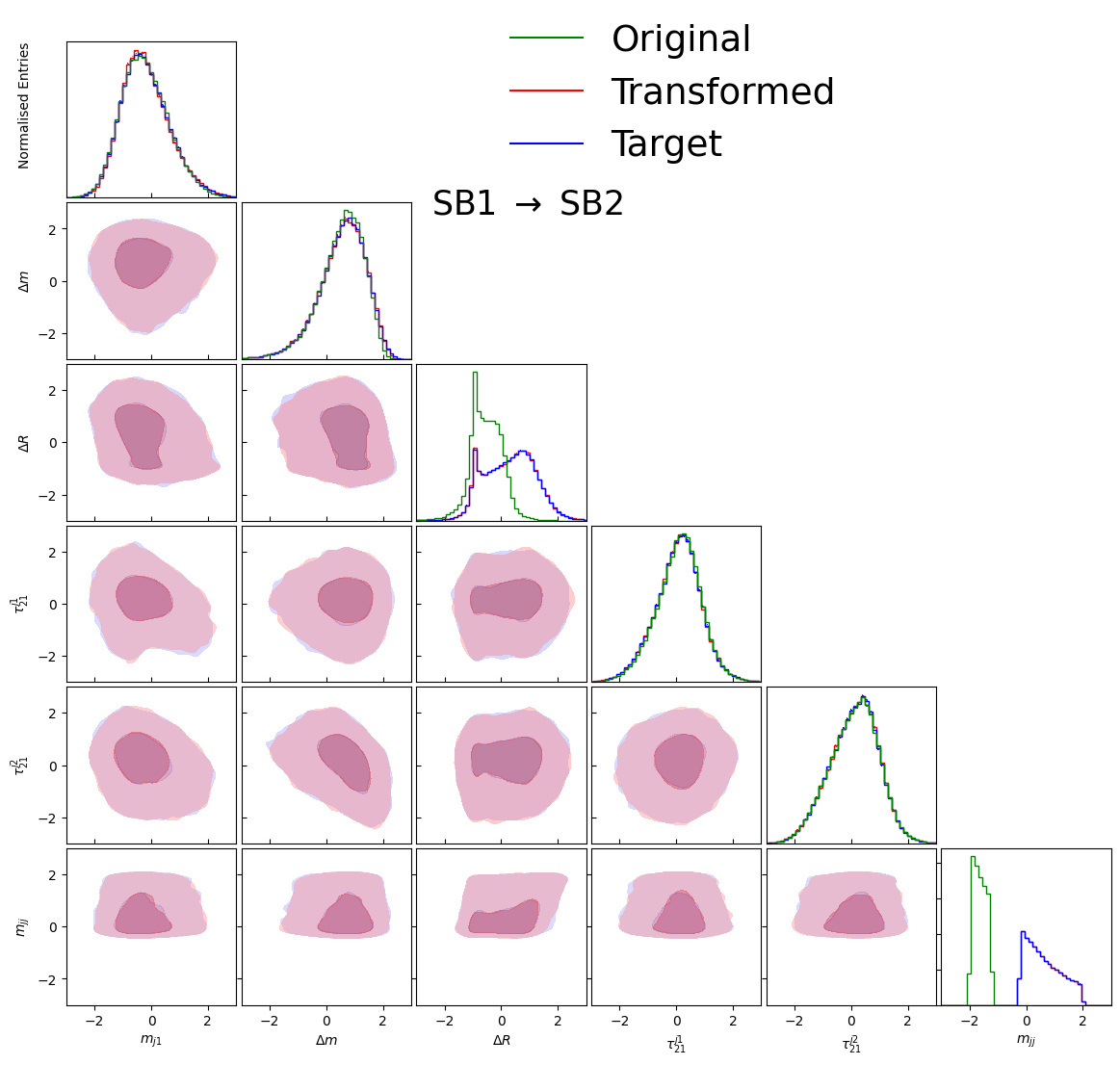}
    \caption{Distributions of events in lower ``origin'' sideband $m_{jj} \in [3.0,\,3.3]$ TeV (green) and in higher ``target'' sideband $m_{jj} \in [3.7,\, 4.6]$ TeV (blue) along with the distribution of events obtained by transporting events from the lower to the higher sideband using TRANSIT (red). The diagonal elements show the marginal distributions of the features, while the off-diagonal elements show the correlations between the features (using KDE contour plots with 16000 points). No signal was added in this run.
    }
    \label{fig:SB1_to_SB2}
\end{figure}

\begin{figure}[H]
    \centering
    \includegraphics[width=\linewidth]{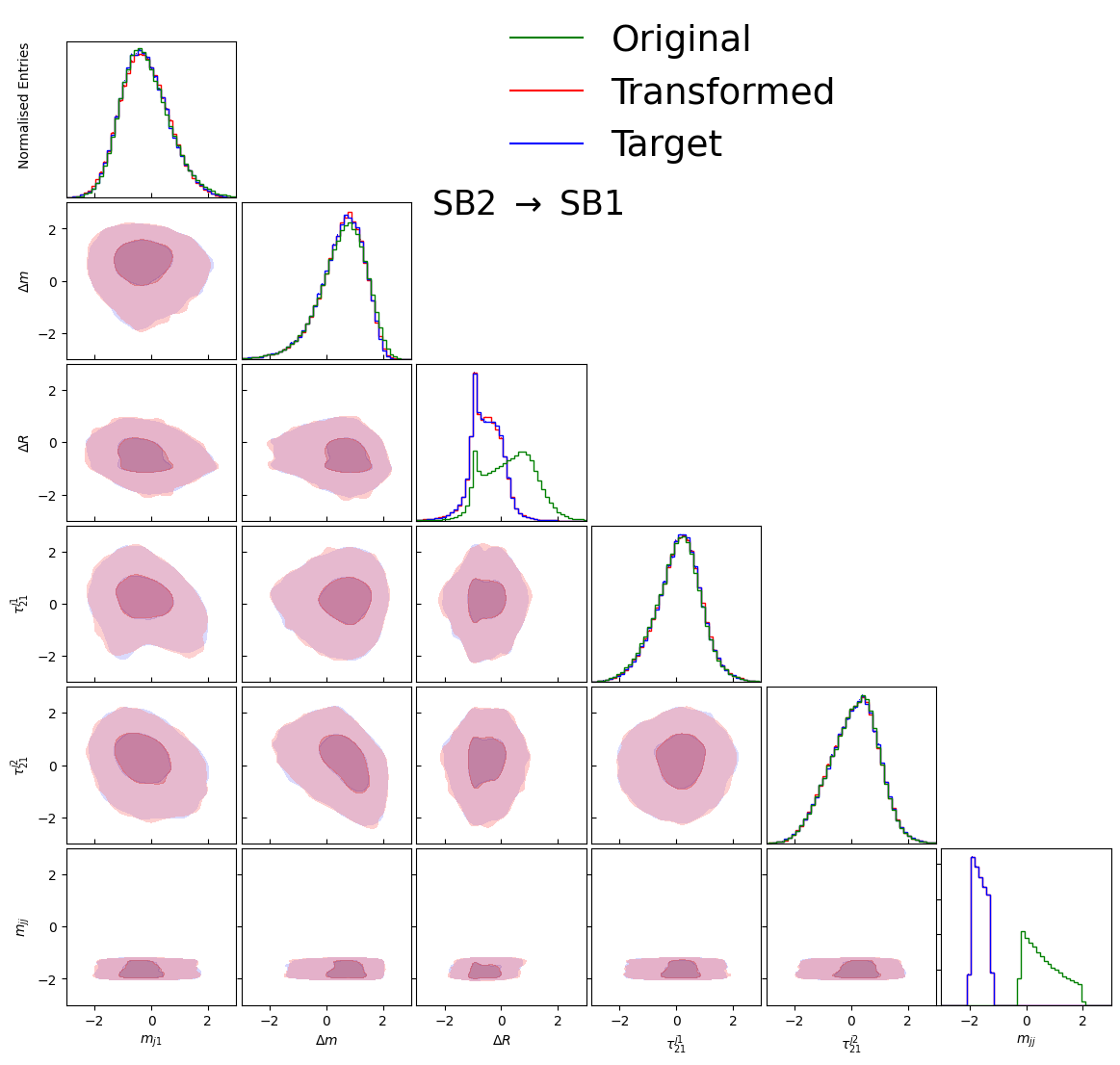}
    \caption{Distributions of events in higher ``origin'' sideband $m_{jj} \in [3.7,\,4.6]$ TeV (green) and lower ``target'' sideband $m_{jj} \in [3.0,\, 4.3]$ TeV (blue) along with the distribution of events obtained by transporting events from the higher to the lower sideband using TRANSIT (red). The diagonal elements show the marginal distributions of the features, while the off-diagonal elements show the correlations between the features (using KDE contour plots with 16000 points). No signal was added in this run.}
    \label{fig:SB2_to_SB1}
\end{figure}

One way to validate the transport quality of the TRANSIT model is to transport events from the first sideband to the second sideband and check that they match the true distribution of events in the second sideband, and vice versa.  
Figs.\,\ref{fig:SB1_to_SB2}, and \ref{fig:SB2_to_SB1} show that the transport is carried out successfully, and both the marginals and the correlations between the variables are well-matched between the transformed and target event sets.  
Such a validation does not require any signal/background labels and can thus be performed on real data.

\section{Transport trajectories}
\label{app:trajectories}

\begin{figure}[h]
    \centering
    \includegraphics[width=0.49\linewidth]{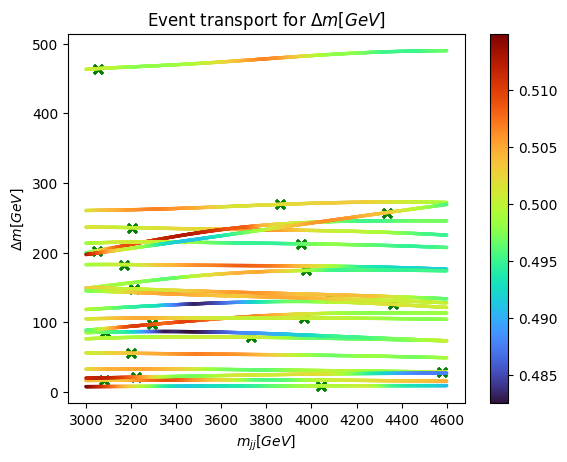}
    \includegraphics[width=0.49\linewidth]{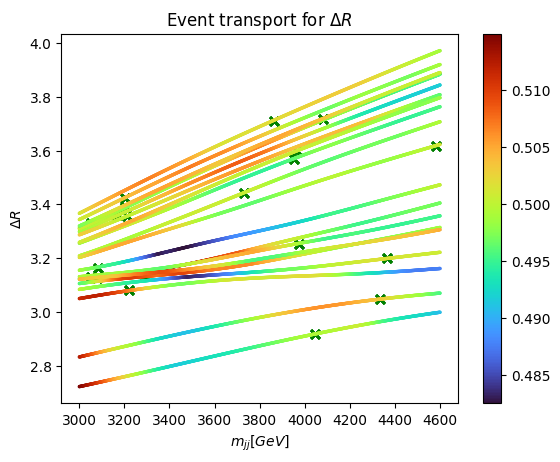}
    \caption{Two-dimensional projections (\(\Delta m\) vs \(m_{jj}\), left, and \(\Delta R\) vs \(m_{jj}\), right) of the transport curves, formed by applying TRANSIT transport to original SB points (green crosses) using an array of different target masses \(\hat{m} = m_{jj}\). The colour map shows the score assigned to each transported point by the adversarial classifier in the TRANSIT model. No signal was added in this run.}
    \label{fig:traj}
\end{figure}

Another data-driven way to ensure the transport quality of the TRANSIT model is to plot the transport curves, shown in Fig.\;\ref{fig:traj}.  
Each curve is created by encoding a point from a sideband region (green cross) into the latent representation of TRANSIT and decoding it using an array of different masses from 3000 GeV to 4600 GeV.  
The distance between each curve and the original point is negligible, showing that the reconstruction loss is well minimised.  
Furthermore, we observe that although some curves are non-linear, all of them are smooth, exhibiting no discontinuities and maintaining moderate curvature at the scale of our problem.  

\begin{figure}[h]
    \centering
    \includegraphics[width=0.49\linewidth]{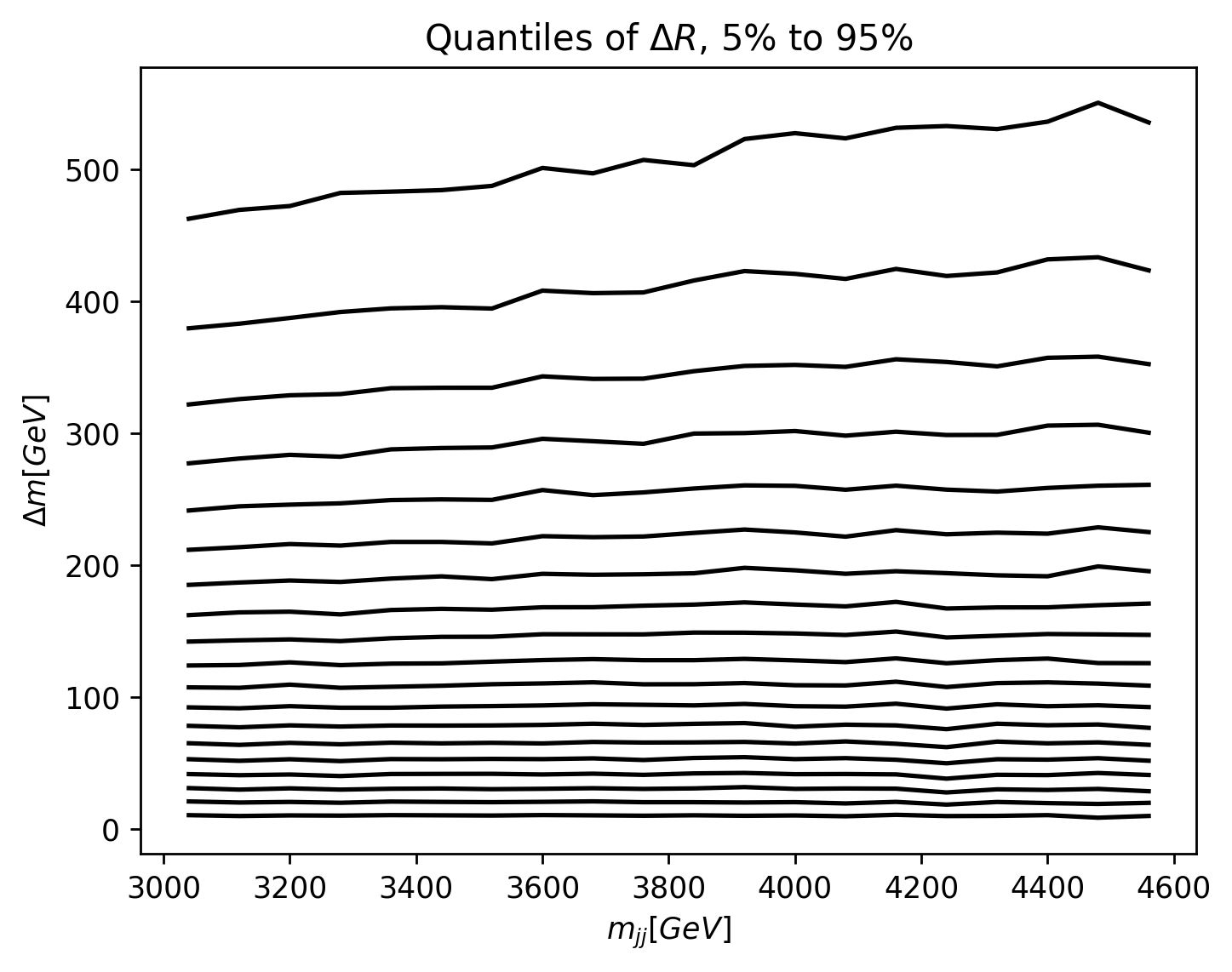}
    \includegraphics[width=0.49\linewidth]{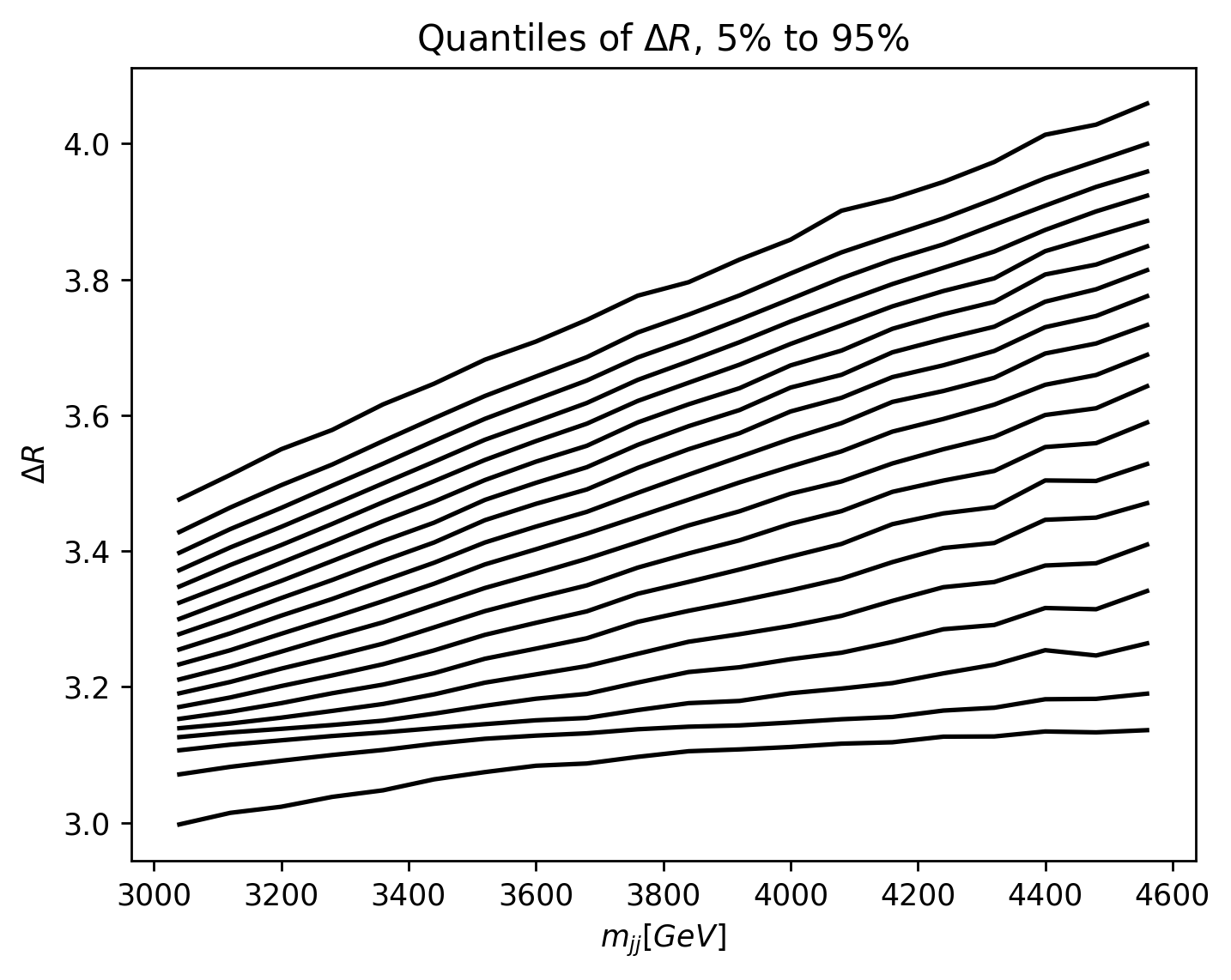}
    \caption{Quantile lines of the conditional distributions $p(\Delta m| m_{jj})$ and $p(\Delta R| m_{jj})$ from 5\% to 95\% with a 5\% increment. The lines are created by finding quantiles in each of the 20 $m_{jj}$ bins with 80 GeV width.}
    \label{fig:quant}
\end{figure}

We can compare these curves to the quantiles of distributions of the same variables shown in Fig.\,\ref{fig:quant}.  
The observable \( \Delta m \) is nearly independent of \( m_{jj} \) in the bulk of the \( \Delta m \) distribution, as evidenced by the flat quantiles.  
\( \Delta m \) only has a small dependence on \( m_{jj} \) in the higher tail of its distribution, corresponding to slightly curved quantiles.  

The TRANSIT transport curves follow the same pattern, providing nearly flat trajectories for the bulk of the \( \Delta m \) distribution, while correctly modelling the tail of the distribution with several curved trajectories at high \( \Delta m \).  
On the other hand, \( \Delta R \) has a strong and partially non-linear correlation with \( m_{jj} \), resulting in non-linear quantiles in Fig.\,\ref{fig:quant}.  
This is reflected in Fig.\,\ref{fig:traj}, where the trajectories exhibit an analogous form, expanding, shifting, and morphing the distribution of \( \Delta R \) for increasing \( m_{jj} \).  
This validates the core idea of TRANSIT: to preserve an uncorrelated variable while smoothly shifting a mass-correlated variable.  

It is important to note that we do not expect the projections of our curves to exactly match the quantiles, except in the simplest cases.  
The reason for this is that order preservation is ill-defined in more than one dimension.  
Thus, for certain pairs of distributions, even optimal transport will yield curves that appear to intersect in some two-dimensional projections, whereas the quantiles of the conditional distribution cannot intersect.  

For each decoded point, we compute the score of the adversarial classifier and display it using a colour map.  
Classifier scores closer to 1 indicate that the adversary identifies the point as background, i.e., our model does not generate enough fake samples in that region.  
Conversely, if the score is closer to 0, the classifier identifies the point on the trajectory as fake.  
In our case, we observe that all scores lie within the \([0.48, 0.52]\) range, meaning that, apart from some minor fluctuations, for any \( m_{jj} \), the generated conditional distribution \( \hat{p}_{\phi,\theta}(\boldsymbol{\hat{X}}|\hat{M}) \) closely matches the true conditional distribution \( p(\boldsymbol{X}|M) \).  
This match is primarily the result of the maximisation of the adversarial discriminator \( \mathcal{L}_{\text{disc}} \) loss by the TM network.  

\section{Example of background sculpting}
\label{app:bkg_sculpting}

\begin{figure}[h]
    \centering
    \includegraphics[width=0.49\linewidth]{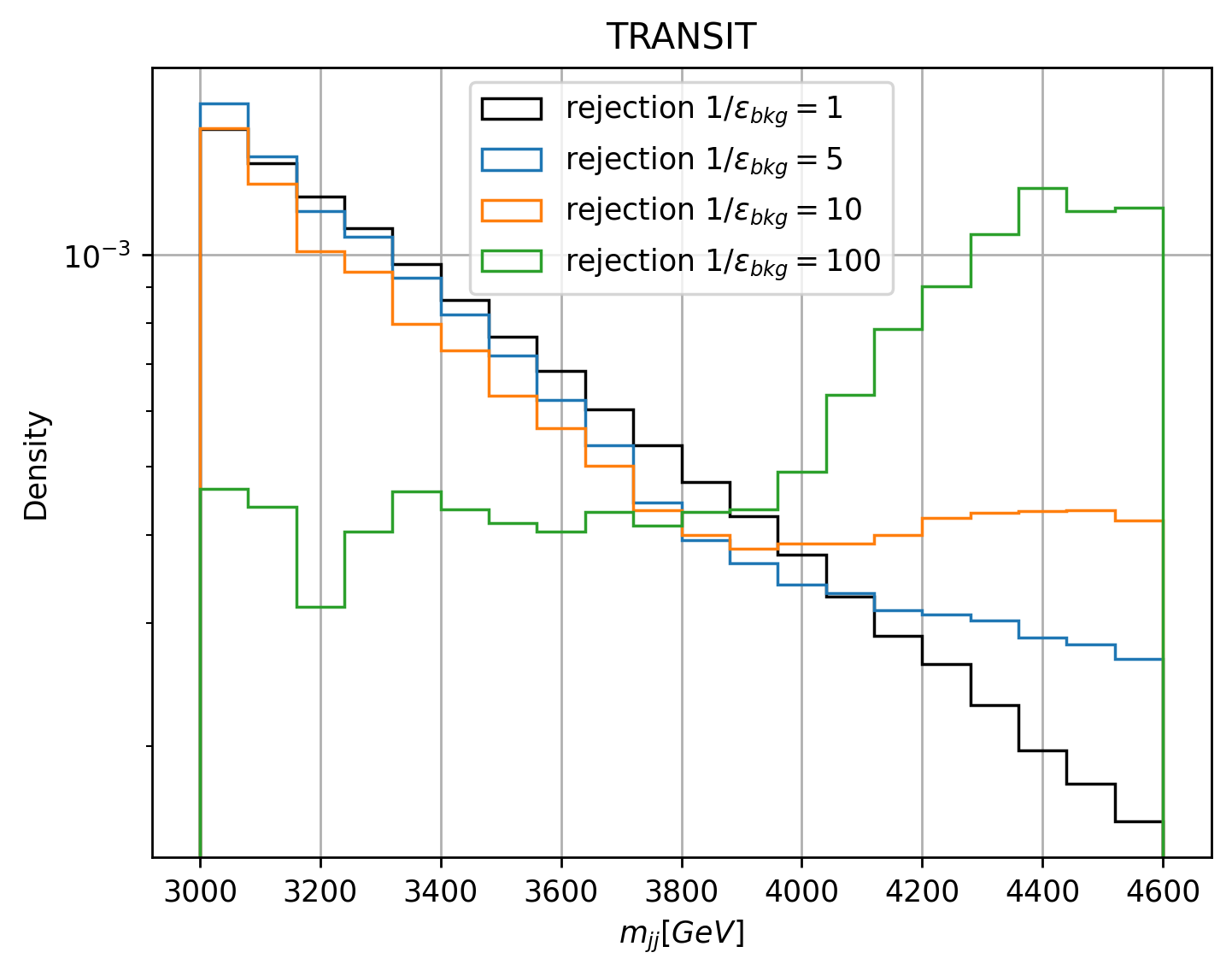}
    \includegraphics[width=0.49\linewidth]{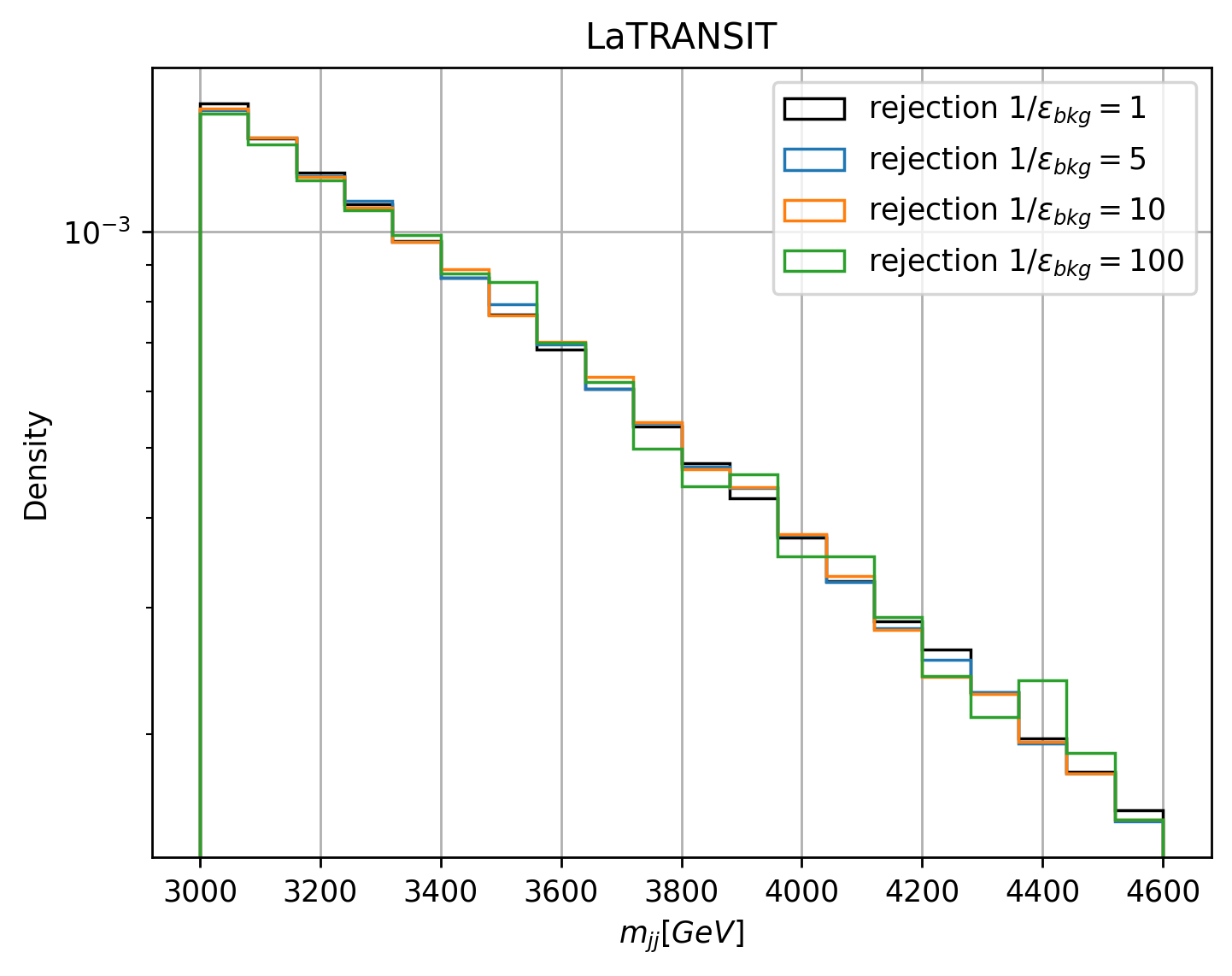}
    \caption{Distribution of the dijet mass $m_{jj}$ after a cut on CWoLa score for TRANSIT (left) and LaTRANSIT (right) methods for one representative model training. No signal was added in this run.}
    \label{fig:sculpting_example}
\end{figure}

Fig.\;\ref{fig:sculpting_example} shows an example of background sculpting after using the score of a CWoLa classifier trained with either the TRANSIT template or the LaTRANSIT latent space representation, both originating from the same TRANSIT network training. 
The original data and TRANSIT template have a mass-dependent \( \Delta R \) observable that induces significant background sculpting. The background distribution deviates further from the original distribution as the rejection threshold increases.
On the other hand, the LaTRANSIT method operates only with mass-independent variables and thus does not exhibit any sculpting beyond the level of statistical fluctuations. 
In general, the shape of the background sculpting depends on the initialisation of both the TRANSIT network and the classifier BDT; here, we provide one representative case.

\newpage

\section{Hyperparameters}
\label{app:hyperparameters}

\begin{table}[H]
\caption{Hyperparameters of the TRANSIT network used for all the results in this publication.}
    \centering
    \begin{tabular}{c c}
        \hline
         Parameter & value \\
         \hline
         \hline
         batch size & 2048 \\
         training epochs & 200 \\
         initial learning rate & $2\times10^{-3}$\\
         learning rate decay on milestone & 0.5 \\
         milestone epochs encoder/decoder &  [30, 100, 150, 175]\\
         milestone epochs discriminator &  [30, 100, 150, 175]\\
         optimiser & AdamW \\
         weight decay &  $1\times10^{-5}$\\
         warmup epochs &  5\\
         \hline
         \hline
         $z$ dimensionality & 8\\
         MLP layers width & 128 \\
         MLP layers per block & 2 \\
         \# residual blocks encoder & 3 \\
         \# residual blocks decoder & 3 \\
         discriminator MLP layer width & [64, 64, 64, 64] \\
         \hline
         \hline
         $w_{rec}$ & 1 \\
         $w_{trans}$ & 0.2 \\
         $w_{cons}$ & 0.1 \\
         \hline
    \end{tabular}
    \label{tab:hyp}
\end{table}

\section{Example for insufficiency of reconstruction and adversarial discriminator losses for round-trip reversibility}
\label{app:conter_reverce}
Imagine a dataset with two features \(x_1\) and \(x_2\), and a conditional feature \(m\), such that the distribution \(p(x_1, x_2 \mid m)\) is uniform in a circle defined by \(x_1^2 + x_2^2 < R^2\), and zero outside of it.
There exists a transformation 
\(
f(\boldsymbol{x}, m, \hat{m}) = \left( r \cos\left(\phi_0 + |m - \hat{m}|\right),\ r \sin\left(\phi_0 + |m - \hat{m}|\right) \right)
\)
where \(r = \sqrt{x_1^2 + x_2^2}\) and \(\phi_0 = \arctan\left(\frac{x_2}{x_1}\right)\), i.e., a rotation by an angle \(\phi = |m - \hat{m}|\), which is a bijection between the distributions \(p(x_1, x_2 \mid m)\) and \(p(x_1, x_2 \mid \hat{m})\) for any fixed $m$ and $\hat{m}$.
As an example, in an architecture defined as \(f(\boldsymbol{x}, m, \hat{m}) = d(e(\boldsymbol{x}, m), \hat{m})\), this can be achieved using an encoder
\(
z = e(\boldsymbol{x}, m) = (r, \phi_0, m)
\)
that maps the input data to a latent space of dimension \(D_{\boldsymbol{z}} = 3 > D_{\boldsymbol{x}}\), and a decoder
\(
d((r, \phi_0, m), \hat{m}) = \left(r \cos(\phi_0 + |m - \hat{m}|),\ r \sin(\phi_0 + |m - \hat{m}|)\right).
\) 
Despite the bijectiviry, the consistency constrain does not hold as $e(f(\boldsymbol{x}, m, \hat{m}), \hat{m})=(r, \phi_0+|m - \hat{m}|, \hat{m})\neq e(\boldsymbol{x}, m) = (r, \phi_0, m)$.

This transformation preserves the conditional density of the data, ensuring that \(p(x_1, x_2 \mid m) = p(x_1, x_2 \mid \hat{m})\). As a result, true samples from \(p(x_1, x_2 \mid \hat{m})\) are indistinguishable from samples transported to \(\hat{m}\), leading the optimal discriminator loss to be \(\ln(4)\). Additionally, for \(m = \hat{m}\), the reconstruction loss is zero, as the transformation reduces to the identity. 
However, this transformation is not round-trip reversible, as
\(
f(f(\boldsymbol{x}, m, \hat{m}), \hat{m}, m) = \left( r \cos\left(\phi_0 + 2|m - \hat{m}|\right),\ r \sin\left(\phi_0 + 2|m - \hat{m}|\right) \right) \neq \boldsymbol{x}
\)
for all $\boldsymbol{x}$ except $\boldsymbol{x}=(0, 0)$.
This example demonstrates that the reconstruction and adversarial losses alone are insufficient to guarantee round-trip reversibility of an arbitrary transport function. Nevertheless, in Subsection~\ref{subsec:transit_model}, we have shown that consistency and reconstruction constraints together are sufficient to ensure round-trip reversibility of an encoder-decoder transport function.

\section{Empirical benefits of the consistency loss}
\label{app:consistency_benefits}

\begin{figure}[h]
    \centering
    \includegraphics[width=0.5\linewidth]{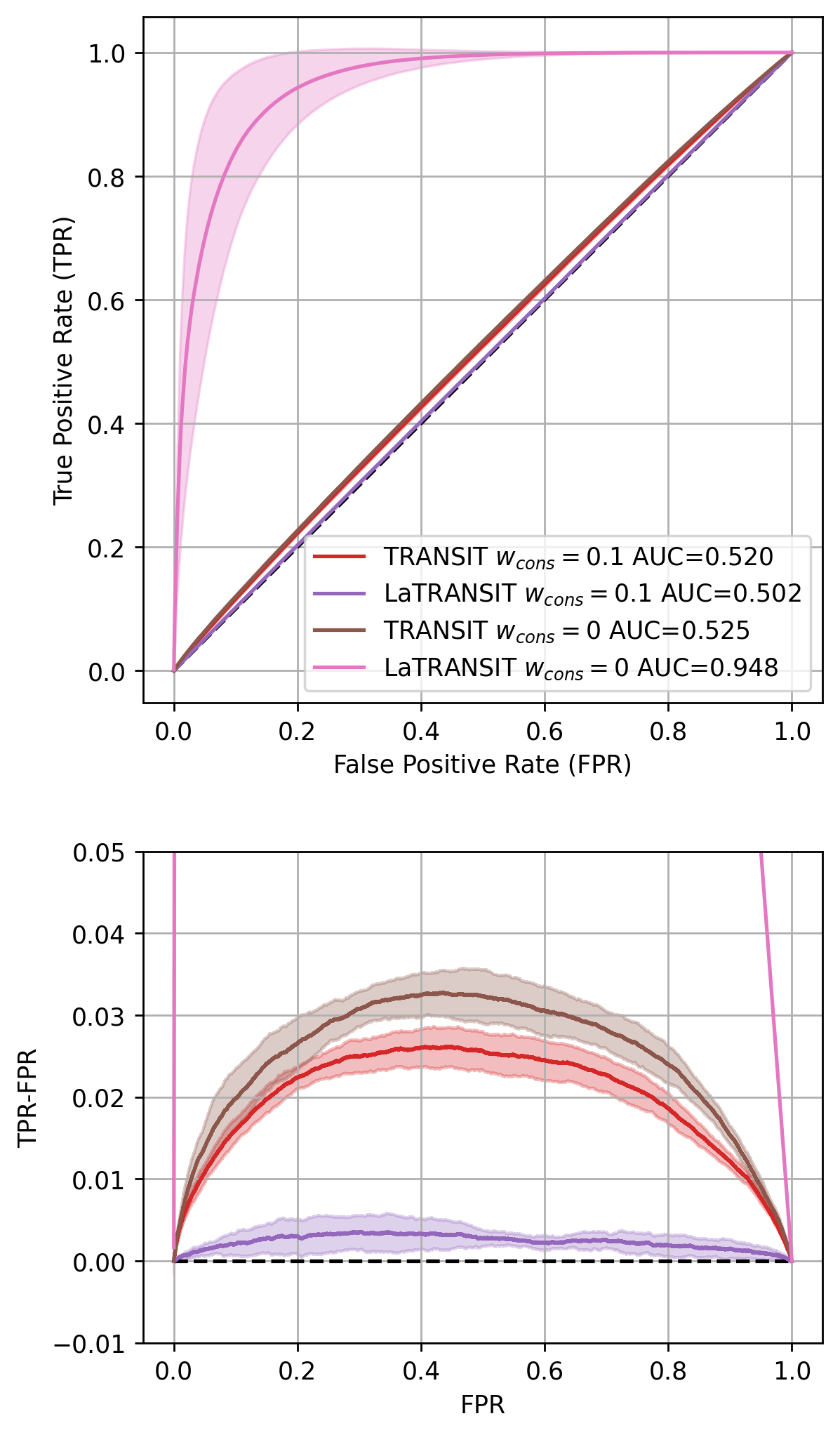}
    \caption{
    ROC curves for a BDT trained to discriminate TRANSIT templates from background SR data and for a BDT trained to discriminate SB latent representations from background SR latent representations in LaTRANSIT. 
    Solid lines and filled regions represent the average and the standard deviation range across 6 TRANSIT network trainings with different initialisation seeds. The comparison is done between TRANSIT with $w_cons=0.1$ (default) and $w_{cons}=0$. No signal was added in these runs.}
    \label{fig:closure_wcons}
\end{figure}

Fig.\,\ref{fig:closure_wcons} extends Fig.\,\ref{fig:closure} from the main text, showing results for the classier closure tests (described in Subsections \ref{subsec:quality} and \ref{subsec:decor}) for TRANSIT model with consistency loss weights of 0.1 (default) and 0 (no consistency loss). 
It is evident that the model without consistency loss exhibits correlations between the mass $m_{jj}$ and the latent features strong enough for the BDT to easily distinguish between the latent representations of background events in the SR and SB regions. 
This demonstrates that the consistency loss is essential for the model to learn a mass-independent latent representation. 
Moreover, we observe a moderate improvement in the TRANSIT template closure when using a non-zero consistency loss weight, further indicating that the consistency loss contributes to better template transport quality.

\newpage